\documentclass[reprint, noshowpacs,onecolumn,superscriptaddress]{revtex4-2}

\usepackage{graphicx}% Include figure files
\usepackage{epstopdf}
\usepackage{comment}
\usepackage{dcolumn}% Align table columns on decimal point
\usepackage{bm}% bold math
\usepackage{amsmath}
\usepackage{amssymb}
\usepackage{braket}
\usepackage[mathlines]{lineno}% Enable numbering of text and display math
\usepackage{units}
\usepackage{upgreek}
\usepackage{chemformula}
\usepackage{orcidlink}
\usepackage{lineno}
\usepackage{subfig}
\renewcommand{\selectlanguage}[1]{}
\definecolor{NewBlue}{rgb}{0, 0, 0.41}
\definecolor{NewRed}{rgb}{0.6, 0.07, 0.07}

\hypersetup{colorlinks,
   linkcolor=NewBlue,
    citecolor=NewBlue,
    urlcolor=NewRed}
\newcommand*{\figref}[2][]{%
  Fig.~\hyperref[{fig:#2}]{%
    \ref*{fig:#2}%
    \ifx\\(#1)\\%
    \else
      (#1)%
    \fi
  }%
}
\begin{document}

\title{Quantum Frequency Conversion of Single Photons from a Tin-Vacancy Center in Diamond}

\author{Julia Maria Brevoord \orcidlink{0000-0002-8801-9616}}
\thanks{These authors contributed equally to this work.}
\affiliation{QuTech and Kavli Institute of Nanoscience, Delft University of Technology, Delft 2628 CJ, The Netherlands}
\author{Jan Fabian Geus \orcidlink{0000-0002-6575-2479}}
\thanks{These authors contributed equally to this work.}
\affiliation{Fraunhofer Institute for Laser Technology ILT, Steinbachstr. 15, D-52074 Aachen, Germany}
\affiliation{RWTH Aachen University LLT - Chair for Laser Technology, Steinbachstr. 15, D-52074 Aachen, Germany}
\author{Tim Turan \orcidlink{0009-0003-9908-7985}}
\affiliation{QuTech and Kavli Institute of Nanoscience, Delft University of Technology, Delft 2628 CJ, The Netherlands}
\author{Miguel Guerrero Romero}
\affiliation{QuTech and Kavli Institute of Nanoscience, Delft University of Technology, Delft 2628 CJ, The Netherlands}
\author{Daniel Bedialauneta Rodríguez \orcidlink{0009-0000-3634-070X}}
\affiliation{QuTech and Kavli Institute of Nanoscience, Delft University of Technology, Delft 2628 CJ, The Netherlands}
\author{Nina Codreanu \orcidlink{0009-0006-6646-8396}}
\affiliation{QuTech and Kavli Institute of Nanoscience, Delft University of Technology, Delft 2628 CJ, The Netherlands}
\author{Alexander Moritz Stramma \orcidlink{0009-0001-5400-9575}}
\affiliation{QuTech and Kavli Institute of Nanoscience, Delft University of Technology, Delft 2628 CJ, The Netherlands}
\author{Ronald Hanson \orcidlink{0000-0001-8938-2137}}
\email{R.Hanson@tudelft.nl}
\affiliation{QuTech and Kavli Institute of Nanoscience, Delft University of Technology, Delft 2628 CJ, The Netherlands}
\author{Florian Elsen \orcidlink{0009-0004-8050-0431}}
\affiliation{Fraunhofer Institute for Laser Technology ILT, Steinbachstr. 15, D-52074 Aachen, Germany}
\affiliation{RWTH Aachen University LLT - Chair for Laser Technology, Steinbachstr. 15, D-52074 Aachen, Germany}
\author{Bernd Jungbluth}
\email{bernd.jungbluth@ilt.fraunhofer.de}
\affiliation{Fraunhofer Institute for Laser Technology ILT, Steinbachstr. 15, D-52074 Aachen, Germany}
\affiliation{RWTH Aachen University LLT - Chair for Laser Technology, Steinbachstr. 15, D-52074 Aachen, Germany}
\begin{abstract} 
Diamond tin-vacancy (SnV) centers are promising candidates for building quantum network nodes. However, their native photon emission at \unit[619]{nm} is incompatible with metropolitan-scale networks operating at low-loss telecom wavelengths. To address this, we demonstrate highly efficient, low-noise quantum frequency conversion (QFC) of \unit[619]{nm} photons to the telecom S-band at \unit[1480]{nm}. The conversion process combines \unit[619]{nm} photons with \unit[1064]{nm} pump light in an actively stabilized cavity containing a bulk monocrystalline potassium titanyl arsenate (KTA) crystal. We achieve an internal (external) conversion efficiency of \unit[(48~$\pm$~3)]{$\%$} (\unit[(28~$\pm$~2)]{$\%$}) and a noise photon rate per wavelength of \unit[(2.2~$\pm$~0.9)]{cts/s/pm}, which is spectrally flat in the investigated frequency range of \unit[40]{GHz}. Furthermore, we demonstrate that the efficiency remains above 80$\%$ of its maximum over a frequency range of \unit[70]{GHz}. Finally, we generate a string of photons from a single waveguide-embedded SnV$^-$ center using a train of excitation pulses and send these through the QFC. After the QFC, we observe a string of telecom photons displaying the SnV lifetime, confirming successful conversion. These results represent a critical step towards metropolitan-scale fiber-based quantum networks using SnV$^-$ centers.
\end{abstract}
\maketitle
%\section{Introduction}
\noindent Quantum networks, consisting of quantum processors interconnected via optical links, have the potential to enable a wide range of applications, including secure communication, distributed quantum computing, networked quantum sensing, and novel tests of fundamental physics~\cite{kimble_quantum_2008, wehner_quantum_2018, ekert_ultimate_2014, jiang_distributed_2007,gottesman_longer-baseline_2012,buhrman_quantum_2001}. Significant progress toward realizing such networks has led to demonstrations of key building blocks of a metropolitan quantum network~\cite{bernien_heralded_2013, hofmann_heralded_2012, moehring_entanglement_2007, ritter_elementary_2012, abobeih_fault-tolerant_2022,bradley_ten-qubit_2019}.
\\
\\
To date, the state-of-the-art multi-node quantum network comprises three nodes~\cite{pompili_realization_2021}, with quantum teleportation demonstrated between non-neighboring nodes~\cite{hermans_qubit_2022}. In terms of on-chip capabilities, Li \textit{et al.} demonstrated an architecture for large-scale heterogeneous integration, calibration, and spectral tuning of spin qubits, and high-fidelity spin state initialization and readout~\cite{li_heterogeneous_2024}. Long-distance heralded quantum entanglement over metropolitan scales has also been achieved, notably using two nitrogen-vacancy (NV) centers in diamond separated by approximately~\unit[10]{km} and connected via \unit[25]{km} of deployed fiber infrastructure~\cite{stolk_metropolitan-scale_2024}. In addition, the entanglement of two nanophotonic quantum memories was demonstrated between nearby nodes connected by a \unit[35]{km} deployed fiber loop~\cite{knaut_entanglement_2024}.
\\
\\
A key element in many quantum networking protocols is the generation and interference of single photons, so-called flying qubits, that are entangled with stationary qubits~\cite{beukers_remote-entanglement_2024}. These photons are typically transmitted through optical fibers, where propagation losses ultimately limit the entanglement generation rate for longer distances. However, most optically active quantum color centers emit photons outside the telecom spectrum~\cite{pompili_realization_2021, daiss_quantum-logic_2021, krutyanskiy_light-matter_2019}, where fiber transmission loss is significantly higher than in the telecom band. Quantum frequency conversion (QFC) offers a viable solution by coherently shifting the photon wavelength into the low-loss telecom regime while preserving its quantum properties~\cite{tchebotareva_entanglement_2019, stolk_telecom-band_2022, knaut_entanglement_2024}.
\\
\\
Among the emerging spin-photon interfaces, the negatively charged tin-vacancy (SnV$^-$) center in diamond has gained attention due to its relatively high quantum efficiency~\cite{iwasaki_tin-vacancy_2017, herrmann_coherent_2024}, large spin-orbit coupling~\cite{karapatzakis_microwave_2024, guo_microwave-based_2023,rosenthal_microwave_2023}, and its inversion symmetry~\cite{thiering_ab_2018}, which allows for nanophotonic integration~\cite{rugar_quantum_2021, arjona_martinez_photonic_2022, pasini_nonlinear_2024, li_heterogeneous_2024, clark_nanoelectromechanical_2024}. The SnV$^-$ center consists of a tin atom positioned at the interstitial site between two adjacent missing carbon atoms in the diamond lattice. In the absence of a magnetic field, the SnV$^-$ has four optical transitions originating from spin-degenerate orbital doublets in the optically ground and excited states. The coherent emission, which is useful for entanglement protocols and quantum applications, is the zero-phonon line (ZPL) transition at~\unit[619]{nm}, also known as the C-transition, which links the lowest branch of the ground and excited states. \\
\\
In this work, we demonstrate low-noise, efficient quantum frequency conversion of single photons emitted from an SnV$^-$ center in diamond. By converting these photons from~\unit[619]{nm} to the telecom S-band at~\unit[1480]{nm}, our approach overcomes a major bottleneck for SnV$^-$-based metropolitan quantum networks with deployed fiber interconnects.

\section{Methods $\&$ Results}
\subsection{Quantum Frequency Converter}
\begin{figure}[!h]%
\centering\includegraphics[width=13.5cm]{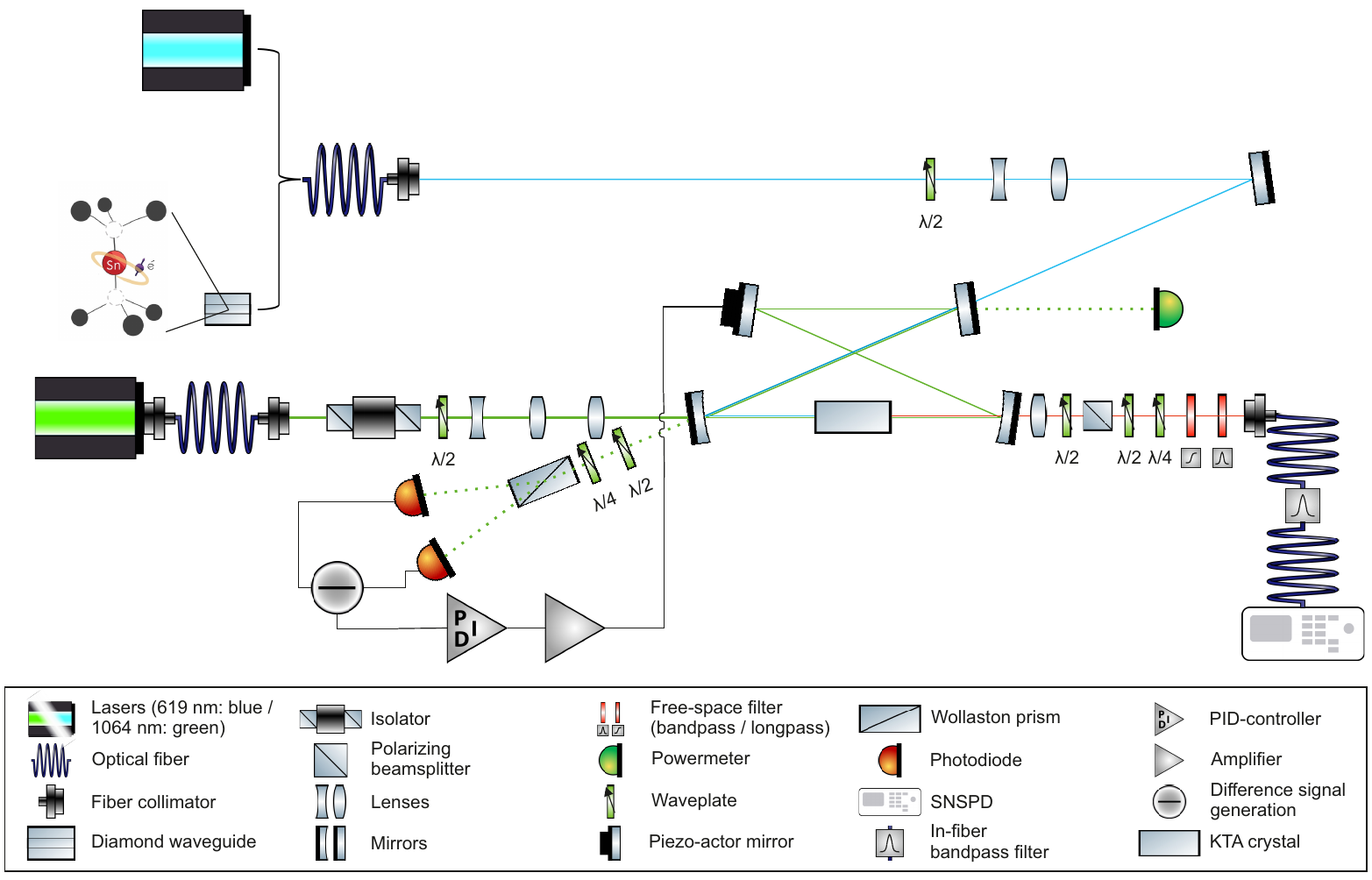}
\caption{\textbf{Schematic of the frequency conversion setup.} Single photons from an SnV$^-$ center are launched into free space and mode-matched via a telescope before entering the KTA crystal. A half-waveplate is used to align the polarization for maximal conversion efficiency. The converted telecom light is filtered using polarization optics, long-pass filters, and a bandpass filter (FWHM = \unit[12]{nm}), and then coupled into a polarization-maintaining fiber. An additional in-fiber bandpass filter (FWHM = \unit[5]{GHz}) precedes detection by a superconducting nanowire single-photon detector (SNSPD). The pump light originates from a continuous wave single-frequency laser, which is beam-shaped to be coupled to the bow-tie cavity containing the KTA crystal. This cavity is locked via the Hänsch-Couillaud technique to maintain resonance. When the internal and external conversion efficiency are measured, the SnV$^-$-emitter setup is replaced with a laser, and the converted light power is measured before the output fiber-collimator and in-fiber, respectively.}
\label{fig: QFC set-up}
\end{figure}

\noindent The QFC setup implemented in this work is adapted from the design reported in~\cite{Geus.2024} and implemented in~\cite{stolk_metropolitan-scale_2024}, modified to enable frequency down-conversion of photons emitted from the zero-phonon line (ZPL) of an SnV$^-$ center at $\uplambda_{\text{SnV}}$~=~\unit[619]{nm}~by a change in the crystal‘s cut orientation. The process relies on three-wave mixing in a bulk monocrystalline potassium
titanyl arsenate (KTA) crystal using a strong intermediate-wavelength pump laser at $\uplambda_{\text{p}}$~=~\unit[1064]{nm} (NKT Adjustik Y10 and Boostik Y10). The output wavelength, $\uplambda_{\text{out}}$, is calculated via energy conservation as,
\begin{equation}
    \uplambda_{\text{out}}~=~(\uplambda_{\text{SnV}}^{-1}-\uplambda_{\text{p}}^{-1})^{-1}~=~\unit[1480]{nm}.
\end{equation} 
Efficient conversion is achieved via birefringent phase matching in the KTA crystal, which is placed within a bow-tie cavity to enhance the pump laser power to the required levels for efficient conversion. The crystal is cut such that the wave vectors of the incident light lie in the plane defined by the crystal's x- and y-axes at a phase matching angle of $\phi~=~40$\textdegree~(measured from x towards y), enabling type-II phase matching, where pump laser (extraordinary) and frequency-converted photons (ordinary) exhibit perpendicular polarization. The nomenclature follows the convention where $n_{x}<n_{y}<n_{z}$. A schematic of the complete setup is shown in Fig.~\ref{fig: QFC set-up}.
\\
The \unit[619]{nm} input is coupled from a polarization-maintaining single-mode fiber to the free-space setup of the QFC via a collimator and mode-shaped with a telescope for optimal overlap with the cavity mode at the focus position located at the crystal's center. The polarization of the \unit[619]{nm} photons is adapted by a half-waveplate to optimize the conversion efficiency.\\
\\
One mirror of the bow-tie cavity is mounted on a piezoelectric actuator for active stabilization using the Hänsch-Couillaud locking scheme~\cite{Hansch.1980}. With an enhancement-factor of about $40$ ($60$), more than \unit[350]{W} of pump laser power is available inside the cavity from the \unit[9]{W} input power (\unit[6]{W} coupled power) delivered by the laser. The enhancement factor and thus conversion efficiency can be increased further by optimizing the impedance matching. To suppress noise from pump scattering and other nonlinear processes (e.g., Raman and spontaneous parametric down-conversion (SPDC)), the converted light is filtered. The filtering setup consists of a combination of \unit[1400]{nm} long-pass filters, polarization filtering, a narrowband bandpass filter (FWHM~=~\unit[12]{nm}), and a narrowband in-fiber Bragg grating (FBG) bandpass filter (TeraXion, FWHM~=~\unit[5]{GHz}).

\subsection{Conversion efficiency and bandwidth}
\noindent When the conversion efficiency is measured, a laser emitting at a wavelength of $\uplambda_{\text{SnV}}=\unit[619]{nm}$ (Toptica DLC TA pro) is used instead of the SnV$^-$ center. The internal (external) conversion efficiency is then calculated by the ratio of converted light power in front of the output fiber-collimator (in-fiber behind the narrowband filter) to the red laser power entering the free-space part of the QFC, corrected for the ratio of photon energies ($\eta_{Q}=1480/619$). To approximate the internal efficiency (i.e., the process's efficiency inside the nonlinear material), the results are corrected for the residual reflectivity of optical coatings of $\unit[8]{\%}$. 
\begin{figure}[h!]
    \centering
    \centering\includegraphics[width=0.725\linewidth]{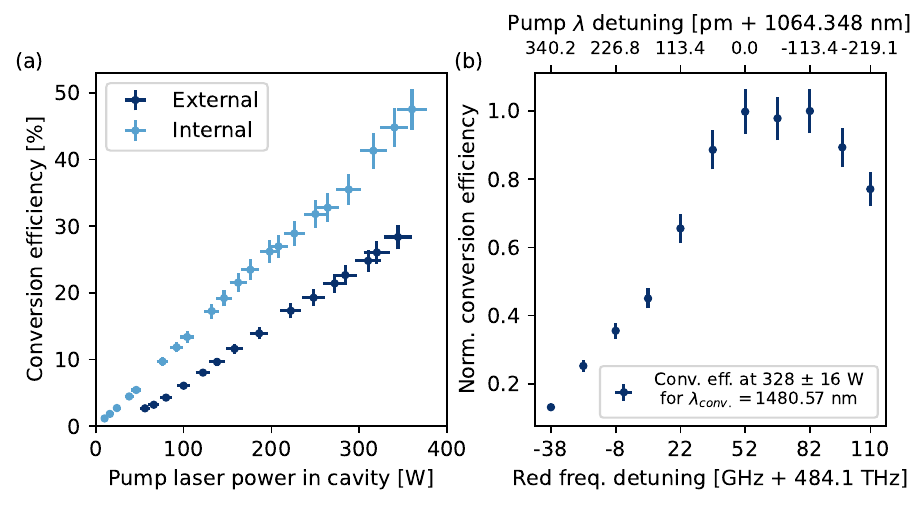}
    \caption{\textbf{Conversion efficiency measured using classical \unit[619]{nm} laser light.} (a) Internal and external conversion efficiencies as a function of circulating pump power. A maximum internal (external) efficiency of \unit[(48~$\pm$~3)]{$\%$} (\unit[(28~$\pm$~2)]{$\%$}) is achieved. The error bars represent the accuracy of the detectors used.  (b) Spectral bandwidth of the QFC setup, measured by varying the red laser input frequency while adjusting the pump wavelength to keep the converted output fixed. Within a \unit[70]{GHz} range, conversion efficiency remains above 80$\%$ of its maximum.}%
    \label{fig: Conversion efficiency}%
\end{figure}%
\\
Fig.~\ref{fig: Conversion efficiency}(a) shows the conversion efficiency as a function of circulating pump power.  At a circulating pump power of~\unit[(360~$\pm$~17)]{W}, an internal efficiency of \unit[(48~$\pm$~3)]{$\%$} is achieved. The error bars are calculated from the accuracy of the power measurements. A maximum external efficiency of \unit[(28~$\pm$~2)]{$\%$} is observed at a pump power of \unit[(344~$\pm$~16)]{W} in the cavity. The conversion efficiency $\eta_{\text{QFC}}$ for a nonlinear material of length $L$ is expected to exhibit the following dependency  on the pump laser power $P$:
\begin{equation}
    \eta_{\text{QFC}}=\eta_{\text{max}} \text{sin}^{2}(L\sqrt{\alpha P}),
\end{equation}
with the maximum conversion efficiency $\eta_{\text{max}}$ and the normalized power efficiency $\alpha$ \cite{Roussev.2004}. The linear increase of the conversion efficiency with pump laser power suggests an unaccounted dependency on either $\eta_{\text{QFC}}$ or $\alpha$. Sweeping the incident pump laser power affects both the thermal load to the nonlinear crystal and the power reaching the photodiodes, which generate the error signal for cavity locking, resulting in a change in the temperature distribution inside the crystal and potentially the spatial properties of the cavity mode.
\\
\\
Next, we probe the spectral bandwidth of the conversion setup, which is relevant in conjunction with the properties of SnV$^-$ center emission. Due to varying local strain in the diamond lattice, different SnV$^-$ centers may exhibit different emission frequencies, indicated by the inhomogeneous distribution. Depending on the SnV$^-$ center creation process, the inhomogeneous distribution can vary from \unit[4]{GHz} to more than \unit[100]{GHz}~\cite{brevoord_large-range_2025, li_heterogeneous_2024,gorlitz_spectroscopic_2020, narita_multiple_2023, bushmakin_two-photon_2024}. In addition, SnV$^-$ centers' resonant frequencies can drift over time. For both scenarios, these frequency differences can be compensated for by simply adjusting the pump laser wavelength and thereby maintaining converted photons with the same frequency. We measure the spectral acceptance bandwidth by scanning the input signal frequency while compensating with the pump wavelength, thus keeping the target wavelength fixed. As shown in Fig.~\ref{fig: Conversion efficiency}(b), the system maintains over $\unit[80]{\%}$ of its maximum efficiency across a \unit[70]{GHz} tuning range.

\subsection{QFC noise characterization}
\begin{figure}[h!]
\centering\includegraphics[width=0.725\linewidth]{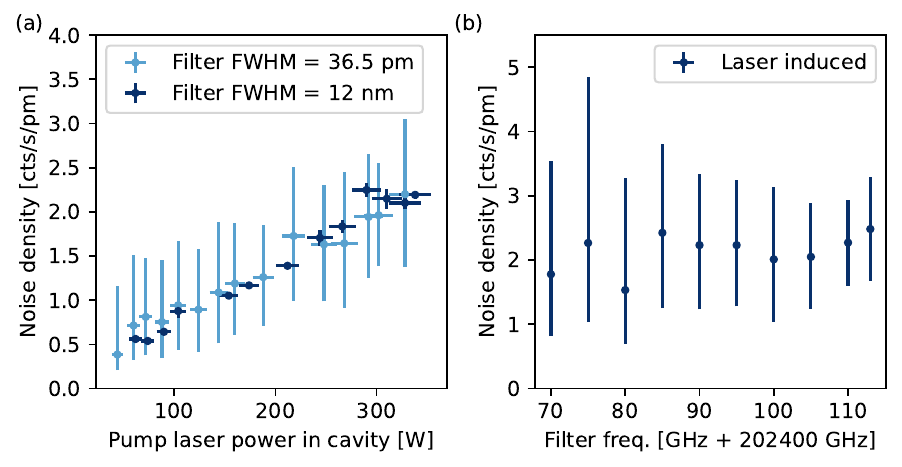}
    \caption{\textbf{Noise characterization.} All measurement data depicted here is corrected for the  SNSPD-efficiency, detector dark-counts, and (if applicable) thermal noise and losses introduced by the tunable narrowband in-fiber filter. A Bayesian approach to error analysis is implemented to prevent a non-physical noise density error bar. (a) Noise density as a function of the pump laser power. The noise increases linearly with the pump laser power up to a value of \unit[(2.2$\pm$0.9)]{cts/s/pm}. The data depict measurements with (without) the narrowband in-fiber filter, resulting in a FWHM of the spectral pass-band of $\unit[36.5]{pm}$ ($\unit[12]{nm}$). (b) Noise density as a function of center frequency of the narrowband in-fiber filter. Within the accuracy of this measurement, spectrally flat noise is measured in the investigated regime.}
    \label{fig: Noise}
\end{figure}

\noindent With the pump laser wavelength being much shorter than the wavelength of frequency-converted photons, Raman-scattered pump photons are not expected to contribute significantly to the generation of noise, leaving SPDC as the expected dominant source of noise photons. To quantify the noise introduced by the QFC process, we measure the count rate (averaged over \unit[20]{s}) on an SNSPD (Quantum Opus, detection efficiency $\eta_{\text{SNSPD}}$= 75$\%$) connected to the QFC setup after the in-fiber narrowband filter while only the pump laser is operating and no red light enters the QFC. All counts in this configuration, therefore, arise either from SPDC of the pump laser light in the crystal or can be attributed to black-body radiation and detector dark counts. The noise density is determined by normalizing the count rate, $N$, to the detector efficiency, $\eta_{\text{SNSPD}}$, and the bandwidth of the spectral filter stack, $\Delta\uplambda_{\text{BP}}$,
\begin{equation}
    \frac{dn}{d\uplambda}=\frac{N}{\eta_{\text{SNSPD}}\Delta\uplambda_{\text{BP}}}.
\end{equation}
When the additional narrowband FBG is used, the data is further corrected for its transmission ($T_{\text{FBG}}=\unit[81]{\%}$) for comparability between the two filter setups.
The results are depicted in Fig.~\ref{fig: Noise}(a). 
We observe a linear increase of the noise density with the pump laser power in the cavity up to a value of \unit[(2.2~$\pm$~0.9)]{cts/s/pm}, which indicates SPDC as the leading noise contribution~\cite{Hong.1985}. Measurements with and without the \unit[5]{GHz} in-fiber bandpass filter yield similar results. Additionally, the center frequency of the FBG is swept at maximum pump laser power (see Fig.~\ref{fig: Noise}(b)), yielding no significant dependency on the measured noise rate as a function of the center wavelength of the filter. Both investigations indicate spectral homogeneity of the SPDC-generated noise.
\subsection{Conversion of single SnV$^-$center photons}
\begin{figure}[h!]
    \centering
    \centering\includegraphics[width=0.725\linewidth]{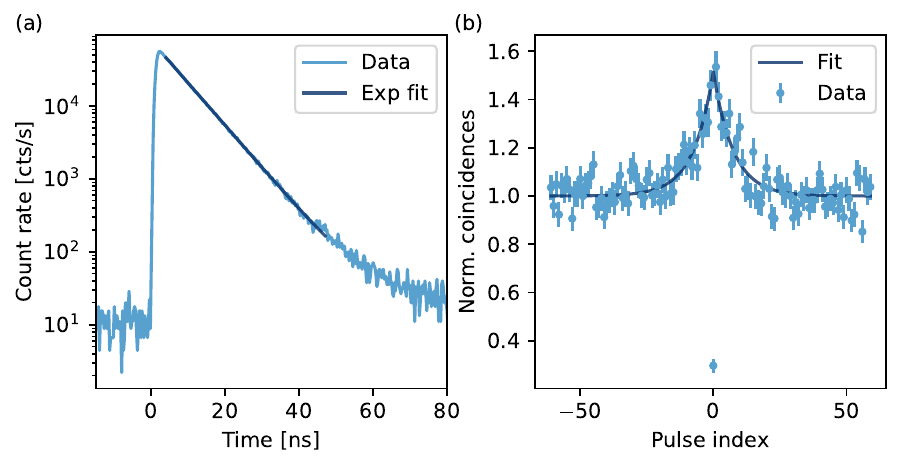}
    \caption{\textbf{Lifetime and second-order correlation function g$^{(2)}(\tau)$ measurements.}  (a) Unconverted lifetime measurement results of the SnV$^-$ center under investigation. The optical decay is fitted with an exponential function f(t)~=~$\text{A} \cdot e^{-t/\tau} + \text{B}$, resulting in an optical lifetime of the SnV$^-$ center of $\tau$ = \unit[(7.47~$\pm$~0.11)]{ns} and A~=~\unit[(54394~$\pm$~1112)]{cts/s} and B~=~\unit[(22~$\pm$~8)]{cts/s}. (b) The second-order correlation function g$^{(2)}(\tau)$ performed in a pulsed fashion. The data is fitted with a bunching model, g$^{(2)}(t)~=~1 + \text{A} \cdot e^{-t/\tau}$, following the procedure of Ref.~\cite{ruf_resonant_2021} resulting in g$^{(2)}$(0)=(0.298$\pm$0.004), without background correction. The fit results include A~=~\unit[(0.51$\pm$0.04)]{} and $\tau$~=~\unit[(7.50$\pm$0.85)]{pulses}.}
    \label{fig: SnV lifetime and g(2)}%
\end{figure}

\begin{figure}[h!]
    \centering
    \includegraphics[width=0.42\linewidth]{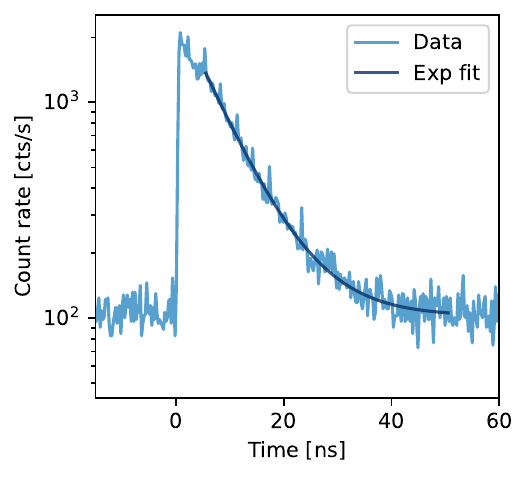}
    \caption{\textbf{SnV$^-$ center emitted converted photons.} The photons from the SnV$^-$ center are converted by the QFC setup and measured with the SNSPDs at a pump laser power of about~\unit[350]{W}. The optical decay is fitted with an exponential function f(t)~=~$\text{A}~\cdot~e^{-t/\tau}~+~\text{B}$, resulting in an optical lifetime of the SnV$^-$ center of $\tau$~=~\unit[(7.58$\pm$~0.14)]{ns} and A~=~\unit[(243~$\pm$~6)$\cdot 10^1$]{cts/s} and B~=~\unit[(102~$\pm$~3)]{cts/s}. The resulting lifetime is similar to the lifetime measured before conversion. This indicates that we have successfully converted photons originating from an SnV$^-$ center to the telecom S-band. We attribute the small periodic peaks to either electronic or optical reflections.}
    \label{fig: Lifetime converted}
\end{figure}

\noindent We finally apply our quantum frequency conversion process to photons emitted by a single SnV$^-$ center. We first describe the generation and characterization of those photons. The SnV$^-$ center studied in this work is embedded in a suspended diamond waveguide device fabricated as described in Ref.~\cite{pasini_nonlinear_2024}. In short, the SnV$^-$ centers are created via tin-ion implantation, followed by acid cleaning and high-temperature vacuum annealing. The waveguide devices are fabricated by a quasi-isotropic undercut etch-based fabrication process. 
\begin{figure}[h!]
\centering\includegraphics[width=11cm]{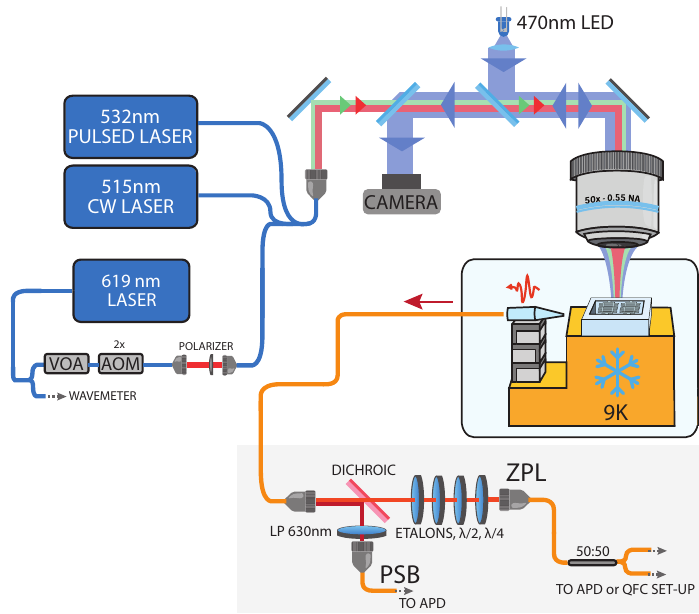}
\caption{SnV$^-$ setup used for the SnV$^-$ experiments. The SnV$^-$ centers are embedded in suspended diamond waveguides, with a tapered fiber positioned via nano-positioners in a lensed configuration in front of the tapered waveguide. This is mounted in a \unit[9]{K} closed-cycle cryostat. Off-resonant excitation is performed in free space using an external objective, with a CW \unit[515]{nm} laser and a pulsed \unit[532]{nm} laser (\unit[1]{MHz} repetition rate). A camera and LED enable sample imaging. Resonant excitation is done with a \unit[619]{nm} Toptica TA-SHG Pro laser. PSB and ZPL light are separated by a dichroic mirror, with an LP\unit[630]{nm} filter removing leaked excitation light. In the ZPL path, narrowband etalons select the C-transition ZPL photons from an SnV$^-$ center. }
\label{fig: SnV set up}
\end{figure}%
\\
Photon collection from the waveguide is achieved using a tapered single-mode optical fiber. The fiber is chemically etched into a conical shape via hydrofluoric acid, producing a lensing effect when positioned in front of the waveguide taper. The diamond sample and fiber are mounted inside a closed-cycle optical cryostat operating at \unit[9]{K}, see Ref. ~\cite{pasini_nonlinear_2024} for details.\\
\\
Resonant excitation of the SnV$^-$ is performed using a Toptica~DLC~TA-SHG~Pro~laser at~\unit[619]{nm} from a free-space objective above the sample. Off-resonant excitation is provided using either a continuous-wave laser (Cobolt 06-MLD, \unit[515]{nm}) or a pulsed laser (NKT Onefive Katana Advanced Laser Diode Systems PiLas, \unit[532]{nm}, \unit[65]{ps} pulse width, \unit[1]{MHz} repetition rate, attenuated to $\sim$\unit[500]{$\upmu$W} average power) for lifetime and second-order correlation measurements.\\
\\
Emission is collected via the tapered fiber and separated into the ZPL and Phonon-Side-Band (PSB) paths using a dichroic mirror, see Fig.~\ref{fig: SnV set up}. The ZPL photons are further filtered using a free space etalon combination, consisting of a~\unit[40]{GHz} and a \unit[4]{GHz} narrowband etalon in series, to suppress photon counts from other SnV$^-$ centers. For the experiments where frequency conversion is not performed, the photons are detected using Laser Component COUNT-T avalanche photodetectors (APD) with a timing-jitter of $\sim$\unit[400]{ps}, connected to a time-correlated single-photon counting module PicoQuant~MultiHarp~150 for time-resolved photon detection. For the conversion experiments, the ZPL photons are routed to the QFC setup via a \unit[35]{m} polarization-maintaining optical fiber.\\
\\
To generate a train of single photons, we apply pulsed off-resonant excitation at \unit[532]{nm} and time-resolve the filtered ZPL emission. The ZPL photons are collected and spectrally filtered to a window of~\unit[4]{GHz}. Fig.~\ref{fig: SnV lifetime and g(2)}(a) shows a histogram of photon detection times relative to the excitation pulse at $t$=0. The counts are normalized to counts per bin over the bin width. The probability of photon detection per excitation pulse is 9.33e-04. An exponential fit indicates that the lifetime is~\unit[(7.47~$\pm$~0.11)]{ns}. The signal-to-noise (SNR) right after the optical pulse exceeds 1000.
\\
\noindent Single-photon statistics are validated via a second-order correlation measurement g$^{(2)}(\tau)$. The emitted photons are split with a 50:50 in-fiber beamsplitter and detected with two detectors. We determine the time difference between detected photons in the different detectors. We integrate the photon counts where the time difference corresponds to a specific pulse index separation. The resulting histogram in Fig.~\ref{fig: SnV lifetime and g(2)}(b) shows a clear dip at zero delay. Note that the data also shows bunching, as the off-resonant excitation leads to an imperfect initialization into the optically bright negatively charged SnV$^-$ state. We obtain g$^{(2)}$(0)=(0.298$\pm$0.004) without any background correction, confirming single-photon emission.
\\
\\
We send the SnV$^-$ center ZPL photons generated by the off-resonant laser pulses to the QFC setup, where we convert the photons to the telecom S-band. The pump laser power in the cavity is maintained at approximately~\unit[350]{W}. Fig.~\ref{fig: Lifetime converted} shows the time-resolved histogram of the converted photons, detected using the SNSPDs. An exponential fit reveals a lifetime of \unit[(7.58$\pm$~0.14)]{ns}, in line with the unconverted value. We determine an SNR of 20 right after the excitation pulse. The signal is reduced to 0.04 of the value before conversion, mainly due to the conversion efficiency (0.28), losses at the free space launching (estimated to be around 0.5), and losses in the \unit[35]{m} visible and \unit[90]{m} telecom optical fiber and connectors. In addition, the noise increased from \unit[22]{cts/s} to \unit[102]{cts/s}, primarily due to pump-laser-induced noise. The combination of a filter with a FWHM of \unit[36.5]{pm} and the pump-laser induced noise of \unit[(2.2~$\pm$~0.9)]{cts/s/pm} determined above, yield ~$\sim$~\unit[73]{cts/s} noise counts. The remaining noise counts are attributed to noise photons in the \unit[90]{m} fiber linking the QFC setup to the SNSPD.
\section{Conclusion}
\noindent We have demonstrated efficient and low-noise frequency down-conversion of single photons emitted from an SnV$^-$ center in diamond, translating its natural wavelength from the visible regime (\unit[619]{nm}) to the telecom S-band (\unit[1480]{nm}). This conversion process is based on the three-wave mixing between a cavity-enhanced pump laser and the single photons in a KTA crystal, with birefringent phase matching enabling high conversion efficiencies. \\
\\
We achieve an internal conversion efficiency of \unit[(48~$\pm$~3)]{$\%$} and an external conversion efficiency of \unit[(28~$\pm$~2)]{$\%$}, while suppressing noise from the scattering process of the high-intensity pump laser through polarization filtering and extensive spectral filtering to \unit[(2.2~$\pm$~0.9)]{cts/s/pm}. The conversion efficiency can be further improved by increasing the pump laser power or improving cavity enhancement.\\
\\
The cavity resonance is tunable to the pump laser frequency, allowing for spectral tunability of the converted photons. This flexibility is crucial for generating indistinguishable photons from multiple quantum emitters with differing emission wavelengths, a key requirement for entanglement generation between different quantum nodes in a quantum network. We demonstrate a conversion bandwidth of \unit[70]{GHz} within which the conversion efficiency remains above 80$\%$ of the maximum, enabling compensating for spectral inhomogeneity between SnV$^-$ centers and spectral drift over time while keeping the target wavelength fixed. \\
\\
Altogether, these results represent an important step toward implementing long-distance, deployed fiber-based quantum networks based on SnV$^-$ centers in diamond. 
\\
\\
\textbf{Funding} % please check
This research was funded by Fraunhofer-Gesellschaft zur Förderung der angewandten Forschung e.V. and by the Dutch Research Council (NWO) through the project “QuTech Part II Applied-oriented research” (project number 601.QT.001) and the Zwaartekracht program Quantum Software Consortium (project no. 024.003.037/3368). We further acknowledge funds from the Dutch Ministry of Economic Affairs and Climate Policy (EZK), as part of the Quantum Delta NL programme, and Holland High Tech through the TKI HTSM (20.0052 PPS) funds. We acknowledge support from the Dutch Research Council (NWO) through the Spinoza prize 2019 (Project No. SPI 63-264).\\
\\
\textbf{Disclosures}
Authors J.F.G., F.E., and B.J. are inventors on International Patent Application \\
WO002023046854A1 and related filings, which are relevant to this work.\\
\\
\textbf{Conflict of Interest}
The authors declare no conflicts of interest.\\
\\
\textbf{Data Availability Statement}
The datasets of this study and the Python software for analysis and plotting are publicly available on 4TU.ResearchData under Ref.~\cite{brevoord_data_2025}.

\bigskip

\bibliography{main}

\begin{thebibliography}{42}%
\makeatletter
\providecommand \@ifxundefined [1]{%
 \@ifx{#1\undefined}
}%
\providecommand \@ifnum [1]{%
 \ifnum #1\expandafter \@firstoftwo
 \else \expandafter \@secondoftwo
 \fi
}%
\providecommand \@ifx [1]{%
 \ifx #1\expandafter \@firstoftwo
 \else \expandafter \@secondoftwo
 \fi
}%
\providecommand \natexlab [1]{#1}%
\providecommand \enquote  [1]{``#1''}%
\providecommand \bibnamefont  [1]{#1}%
\providecommand \bibfnamefont [1]{#1}%
\providecommand \citenamefont [1]{#1}%
\providecommand \href@noop [0]{\@secondoftwo}%
\providecommand \href [0]{\begingroup \@sanitize@url \@href}%
\providecommand \@href[1]{\@@startlink{#1}\@@href}%
\providecommand \@@href[1]{\endgroup#1\@@endlink}%
\providecommand \@sanitize@url [0]{\catcode `\\12\catcode `\$12\catcode `\&12\catcode `\#12\catcode `\^12\catcode `\_12\catcode `\%12\relax}%
\providecommand \@@startlink[1]{}%
\providecommand \@@endlink[0]{}%
\providecommand \url  [0]{\begingroup\@sanitize@url \@url }%
\providecommand \@url [1]{\endgroup\@href {#1}{\urlprefix }}%
\providecommand \urlprefix  [0]{URL }%
\providecommand \Eprint [0]{\href }%
\providecommand \doibase [0]{https://doi.org/}%
\providecommand \selectlanguage [0]{\@gobble}%
\providecommand \bibinfo  [0]{\@secondoftwo}%
\providecommand \bibfield  [0]{\@secondoftwo}%
\providecommand \translation [1]{[#1]}%
\providecommand \BibitemOpen [0]{}%
\providecommand \bibitemStop [0]{}%
\providecommand \bibitemNoStop [0]{.\EOS\space}%
\providecommand \EOS [0]{\spacefactor3000\relax}%
\providecommand \BibitemShut  [1]{\csname bibitem#1\endcsname}%
\let\auto@bib@innerbib\@empty
%</preamble>
\bibitem [{\citenamefont {Kimble}(2008)}]{kimble_quantum_2008}%
  \BibitemOpen
  \bibfield  {author} {\bibinfo {author} {\bibfnamefont {H.~J.}\ \bibnamefont {Kimble}},\ }\bibfield  {title} {\bibinfo {title} {The quantum internet},\ }\href {https://doi.org/10.1038/nature07127} {\bibfield  {journal} {\bibinfo  {journal} {Nature}\ }\textbf {\bibinfo {volume} {453}},\ \bibinfo {pages} {1023} (\bibinfo {year} {2008})}\BibitemShut {NoStop}%
\bibitem [{\citenamefont {Wehner}\ \emph {et~al.}(2018)\citenamefont {Wehner}, \citenamefont {Elkouss},\ and\ \citenamefont {Hanson}}]{wehner_quantum_2018}%
  \BibitemOpen
  \bibfield  {author} {\bibinfo {author} {\bibfnamefont {S.}~\bibnamefont {Wehner}}, \bibinfo {author} {\bibfnamefont {D.}~\bibnamefont {Elkouss}},\ and\ \bibinfo {author} {\bibfnamefont {R.}~\bibnamefont {Hanson}},\ }\bibfield  {title} {\bibinfo {title} {Quantum internet: {A} vision for the road ahead},\ }\href {https://www.science.org/doi/10.1126/science.aam9288} {\bibfield  {journal} {\bibinfo  {journal} {Science}\ }\textbf {\bibinfo {volume} {362}},\ \bibinfo {pages} {6412} (\bibinfo {year} {2018})}\BibitemShut {NoStop}%
\bibitem [{\citenamefont {Ekert}\ and\ \citenamefont {Renner}(2014)}]{ekert_ultimate_2014}%
  \BibitemOpen
  \bibfield  {author} {\bibinfo {author} {\bibfnamefont {A.}~\bibnamefont {Ekert}}\ and\ \bibinfo {author} {\bibfnamefont {R.}~\bibnamefont {Renner}},\ }\bibfield  {title} {{\selectlanguage {en}\bibinfo {title} {The ultimate physical limits of privacy}},\ }\href {https://doi.org/10.1038/nature13132} {\bibfield  {journal} {\bibinfo  {journal} {Nature}\ }\textbf {\bibinfo {volume} {507}},\ \bibinfo {pages} {443} (\bibinfo {year} {2014})}\BibitemShut {NoStop}%
\bibitem [{\citenamefont {Jiang}\ \emph {et~al.}(2007)\citenamefont {Jiang}, \citenamefont {Taylor}, \citenamefont {Sørensen},\ and\ \citenamefont {Lukin}}]{jiang_distributed_2007}%
  \BibitemOpen
  \bibfield  {author} {\bibinfo {author} {\bibfnamefont {L.}~\bibnamefont {Jiang}}, \bibinfo {author} {\bibfnamefont {J.~M.}\ \bibnamefont {Taylor}}, \bibinfo {author} {\bibfnamefont {A.~S.}\ \bibnamefont {Sørensen}},\ and\ \bibinfo {author} {\bibfnamefont {M.~D.}\ \bibnamefont {Lukin}},\ }\bibfield  {title} {{\selectlanguage {en}\bibinfo {title} {Distributed quantum computation based on small quantum registers}},\ }\href {https://doi.org/10.1103/PhysRevA.76.062323} {\bibfield  {journal} {\bibinfo  {journal} {Phys. Rev. A}\ }\textbf {\bibinfo {volume} {76}},\ \bibinfo {pages} {062323} (\bibinfo {year} {2007})}\BibitemShut {NoStop}%
\bibitem [{\citenamefont {Gottesman}\ \emph {et~al.}(2012)\citenamefont {Gottesman}, \citenamefont {Jennewein},\ and\ \citenamefont {Croke}}]{gottesman_longer-baseline_2012}%
  \BibitemOpen
  \bibfield  {author} {\bibinfo {author} {\bibfnamefont {D.}~\bibnamefont {Gottesman}}, \bibinfo {author} {\bibfnamefont {T.}~\bibnamefont {Jennewein}},\ and\ \bibinfo {author} {\bibfnamefont {S.}~\bibnamefont {Croke}},\ }\bibfield  {title} {{\selectlanguage {en}\bibinfo {title} {Longer-{Baseline} {Telescopes} {Using} {Quantum} {Repeaters}}},\ }\href {https://doi.org/10.1103/PhysRevLett.109.070503} {\bibfield  {journal} {\bibinfo  {journal} {Phys. Rev. Lett.}\ }\textbf {\bibinfo {volume} {109}},\ \bibinfo {pages} {070503} (\bibinfo {year} {2012})}\BibitemShut {NoStop}%
\bibitem [{\citenamefont {Buhrman}\ \emph {et~al.}(2001)\citenamefont {Buhrman}, \citenamefont {Cleve}, \citenamefont {Watrous},\ and\ \citenamefont {Wolf}}]{buhrman_quantum_2001}%
  \BibitemOpen
  \bibfield  {author} {\bibinfo {author} {\bibfnamefont {H.}~\bibnamefont {Buhrman}}, \bibinfo {author} {\bibfnamefont {R.}~\bibnamefont {Cleve}}, \bibinfo {author} {\bibfnamefont {J.}~\bibnamefont {Watrous}},\ and\ \bibinfo {author} {\bibfnamefont {R.~d.}\ \bibnamefont {Wolf}},\ }\bibfield  {title} {{\selectlanguage {en}\bibinfo {title} {Quantum {Fingerprinting}}},\ }\href {https://doi.org/10.1103/PhysRevLett.87.167902} {\bibfield  {journal} {\bibinfo  {journal} {Phys. Rev. Lett.}\ }\textbf {\bibinfo {volume} {87}},\ \bibinfo {pages} {167902} (\bibinfo {year} {2001})}\BibitemShut {NoStop}%
\bibitem [{\citenamefont {Bernien}\ \emph {et~al.}(2013)\citenamefont {Bernien}, \citenamefont {Hensen}, \citenamefont {Pfaff}, \citenamefont {Koolstra}, \citenamefont {Blok}, \citenamefont {Robledo}, \citenamefont {Taminiau}, \citenamefont {Markham}, \citenamefont {Twitchen}, \citenamefont {Childress},\ and\ \citenamefont {Hanson}}]{bernien_heralded_2013}%
  \BibitemOpen
  \bibfield  {author} {\bibinfo {author} {\bibfnamefont {H.}~\bibnamefont {Bernien}}, \bibinfo {author} {\bibfnamefont {B.}~\bibnamefont {Hensen}}, \bibinfo {author} {\bibfnamefont {W.}~\bibnamefont {Pfaff}}, \bibinfo {author} {\bibfnamefont {G.}~\bibnamefont {Koolstra}}, \bibinfo {author} {\bibfnamefont {M.~S.}\ \bibnamefont {Blok}}, \bibinfo {author} {\bibfnamefont {L.}~\bibnamefont {Robledo}}, \bibinfo {author} {\bibfnamefont {T.~H.}\ \bibnamefont {Taminiau}}, \bibinfo {author} {\bibfnamefont {M.}~\bibnamefont {Markham}}, \bibinfo {author} {\bibfnamefont {D.~J.}\ \bibnamefont {Twitchen}}, \bibinfo {author} {\bibfnamefont {L.}~\bibnamefont {Childress}},\ and\ \bibinfo {author} {\bibfnamefont {R.}~\bibnamefont {Hanson}},\ }\bibfield  {title} {\bibinfo {title} {Heralded entanglement between solid-state qubits separated by three metres},\ }\href {https://doi.org/10.1038/nature12016} {\bibfield  {journal} {\bibinfo  {journal} {Nature}\ }\textbf {\bibinfo {volume} {497}},\ \bibinfo {pages} {86} (\bibinfo {year}
  {2013})}\BibitemShut {NoStop}%
\bibitem [{\citenamefont {Hofmann}\ \emph {et~al.}(2012)\citenamefont {Hofmann}, \citenamefont {Krug}, \citenamefont {Ortegel}, \citenamefont {Gérard}, \citenamefont {Weber}, \citenamefont {Rosenfeld},\ and\ \citenamefont {Weinfurter}}]{hofmann_heralded_2012}%
  \BibitemOpen
  \bibfield  {author} {\bibinfo {author} {\bibfnamefont {J.}~\bibnamefont {Hofmann}}, \bibinfo {author} {\bibfnamefont {M.}~\bibnamefont {Krug}}, \bibinfo {author} {\bibfnamefont {N.}~\bibnamefont {Ortegel}}, \bibinfo {author} {\bibfnamefont {L.}~\bibnamefont {Gérard}}, \bibinfo {author} {\bibfnamefont {M.}~\bibnamefont {Weber}}, \bibinfo {author} {\bibfnamefont {W.}~\bibnamefont {Rosenfeld}},\ and\ \bibinfo {author} {\bibfnamefont {H.}~\bibnamefont {Weinfurter}},\ }\bibfield  {title} {\bibinfo {title} {Heralded {Entanglement} {Between} {Widely} {Separated} {Atoms}},\ }\href {https://doi.org/10.1126/science.1221856} {\bibfield  {journal} {\bibinfo  {journal} {Science}\ }\textbf {\bibinfo {volume} {337}},\ \bibinfo {pages} {6090} (\bibinfo {year} {2012})}\BibitemShut {NoStop}%
\bibitem [{\citenamefont {Moehring}\ \emph {et~al.}(2007)\citenamefont {Moehring}, \citenamefont {Maunz}, \citenamefont {Olmschenk}, \citenamefont {Younge}, \citenamefont {Matsukevich}, \citenamefont {Duan},\ and\ \citenamefont {Monroe}}]{moehring_entanglement_2007}%
  \BibitemOpen
  \bibfield  {author} {\bibinfo {author} {\bibfnamefont {D.~L.}\ \bibnamefont {Moehring}}, \bibinfo {author} {\bibfnamefont {P.}~\bibnamefont {Maunz}}, \bibinfo {author} {\bibfnamefont {S.}~\bibnamefont {Olmschenk}}, \bibinfo {author} {\bibfnamefont {K.~C.}\ \bibnamefont {Younge}}, \bibinfo {author} {\bibfnamefont {D.~N.}\ \bibnamefont {Matsukevich}}, \bibinfo {author} {\bibfnamefont {L.-M.}\ \bibnamefont {Duan}},\ and\ \bibinfo {author} {\bibfnamefont {C.}~\bibnamefont {Monroe}},\ }\bibfield  {title} {\bibinfo {title} {Entanglement of {Single}-{Atom} {Quantum} {Bits} at a {Distance}},\ }\href {https://doi.org/10.1038/nature06118} {\bibfield  {journal} {\bibinfo  {journal} {Nature}\ }\textbf {\bibinfo {volume} {449}},\ \bibinfo {pages} {7158} (\bibinfo {year} {2007})}\BibitemShut {NoStop}%
\bibitem [{\citenamefont {Ritter}\ \emph {et~al.}(2012)\citenamefont {Ritter}, \citenamefont {Nölleke}, \citenamefont {Hahn}, \citenamefont {Reiserer}, \citenamefont {Neuzner}, \citenamefont {Uphoff}, \citenamefont {Mücke}, \citenamefont {Figueroa}, \citenamefont {Bochmann},\ and\ \citenamefont {Rempe}}]{ritter_elementary_2012}%
  \BibitemOpen
  \bibfield  {author} {\bibinfo {author} {\bibfnamefont {S.}~\bibnamefont {Ritter}}, \bibinfo {author} {\bibfnamefont {C.}~\bibnamefont {Nölleke}}, \bibinfo {author} {\bibfnamefont {C.}~\bibnamefont {Hahn}}, \bibinfo {author} {\bibfnamefont {A.}~\bibnamefont {Reiserer}}, \bibinfo {author} {\bibfnamefont {A.}~\bibnamefont {Neuzner}}, \bibinfo {author} {\bibfnamefont {M.}~\bibnamefont {Uphoff}}, \bibinfo {author} {\bibfnamefont {M.}~\bibnamefont {Mücke}}, \bibinfo {author} {\bibfnamefont {E.}~\bibnamefont {Figueroa}}, \bibinfo {author} {\bibfnamefont {J.}~\bibnamefont {Bochmann}},\ and\ \bibinfo {author} {\bibfnamefont {G.}~\bibnamefont {Rempe}},\ }\bibfield  {title} {{\selectlanguage {en}\bibinfo {title} {An elementary quantum network of single atoms in optical cavities}},\ }\href {https://doi.org/10.1038/nature11023} {\bibfield  {journal} {\bibinfo  {journal} {Nature}\ }\textbf {\bibinfo {volume} {484}},\ \bibinfo {pages} {195} (\bibinfo {year} {2012})}\BibitemShut {NoStop}%
\bibitem [{\citenamefont {Abobeih}\ \emph {et~al.}(2022)\citenamefont {Abobeih}, \citenamefont {Wang}, \citenamefont {Randall}, \citenamefont {Loenen}, \citenamefont {Bradley}, \citenamefont {Markham}, \citenamefont {Twitchen}, \citenamefont {Terhal},\ and\ \citenamefont {Taminiau}}]{abobeih_fault-tolerant_2022}%
  \BibitemOpen
  \bibfield  {author} {\bibinfo {author} {\bibfnamefont {M.~H.}\ \bibnamefont {Abobeih}}, \bibinfo {author} {\bibfnamefont {Y.}~\bibnamefont {Wang}}, \bibinfo {author} {\bibfnamefont {J.}~\bibnamefont {Randall}}, \bibinfo {author} {\bibfnamefont {S.~J.~H.}\ \bibnamefont {Loenen}}, \bibinfo {author} {\bibfnamefont {C.~E.}\ \bibnamefont {Bradley}}, \bibinfo {author} {\bibfnamefont {M.}~\bibnamefont {Markham}}, \bibinfo {author} {\bibfnamefont {D.~J.}\ \bibnamefont {Twitchen}}, \bibinfo {author} {\bibfnamefont {B.~M.}\ \bibnamefont {Terhal}},\ and\ \bibinfo {author} {\bibfnamefont {T.~H.}\ \bibnamefont {Taminiau}},\ }\bibfield  {title} {{\selectlanguage {en}\bibinfo {title} {Fault-tolerant operation of a logical qubit in a diamond quantum processor}},\ }\href {https://doi.org/10.1038/s41586-022-04819-6} {\bibfield  {journal} {\bibinfo  {journal} {Nature}\ }\textbf {\bibinfo {volume} {606}},\ \bibinfo {pages} {884} (\bibinfo {year} {2022})}\BibitemShut {NoStop}%
\bibitem [{\citenamefont {Bradley}\ \emph {et~al.}(2019)\citenamefont {Bradley}, \citenamefont {Randall}, \citenamefont {Abobeih}, \citenamefont {Berrevoets}, \citenamefont {Degen}, \citenamefont {Bakker}, \citenamefont {Markham}, \citenamefont {Twitchen},\ and\ \citenamefont {Taminiau}}]{bradley_ten-qubit_2019}%
  \BibitemOpen
  \bibfield  {author} {\bibinfo {author} {\bibfnamefont {C.}~\bibnamefont {Bradley}}, \bibinfo {author} {\bibfnamefont {J.}~\bibnamefont {Randall}}, \bibinfo {author} {\bibfnamefont {M.}~\bibnamefont {Abobeih}}, \bibinfo {author} {\bibfnamefont {R.}~\bibnamefont {Berrevoets}}, \bibinfo {author} {\bibfnamefont {M.}~\bibnamefont {Degen}}, \bibinfo {author} {\bibfnamefont {M.}~\bibnamefont {Bakker}}, \bibinfo {author} {\bibfnamefont {M.}~\bibnamefont {Markham}}, \bibinfo {author} {\bibfnamefont {D.}~\bibnamefont {Twitchen}},\ and\ \bibinfo {author} {\bibfnamefont {T.}~\bibnamefont {Taminiau}},\ }\bibfield  {title} {{\selectlanguage {english}\bibinfo {title} {A {Ten}-{Qubit} {Solid}-{State} {Spin} {Register} with {Quantum} {Memory} up to {One} {Minute}}},\ }\href {https://doi.org/10.1103/PhysRevX.9.031045} {\bibfield  {journal} {\bibinfo  {journal} {Phys. Rev. X}\ }\textbf {\bibinfo {volume} {9}},\ \bibinfo {pages} {3} (\bibinfo {year} {2019})}\BibitemShut {NoStop}%
\bibitem [{\citenamefont {Pompili}\ \emph {et~al.}(2021)\citenamefont {Pompili}, \citenamefont {Hermans}, \citenamefont {Baier}, \citenamefont {Beukers}, \citenamefont {Humphreys}, \citenamefont {Schouten}, \citenamefont {Vermeulen}, \citenamefont {Tiggelman}, \citenamefont {Martins}, \citenamefont {Dirkse}, \citenamefont {Wehner},\ and\ \citenamefont {Hanson}}]{pompili_realization_2021}%
  \BibitemOpen
  \bibfield  {author} {\bibinfo {author} {\bibfnamefont {M.}~\bibnamefont {Pompili}}, \bibinfo {author} {\bibfnamefont {S.~L.~N.}\ \bibnamefont {Hermans}}, \bibinfo {author} {\bibfnamefont {S.}~\bibnamefont {Baier}}, \bibinfo {author} {\bibfnamefont {H.~K.~C.}\ \bibnamefont {Beukers}}, \bibinfo {author} {\bibfnamefont {P.~C.}\ \bibnamefont {Humphreys}}, \bibinfo {author} {\bibfnamefont {R.~N.}\ \bibnamefont {Schouten}}, \bibinfo {author} {\bibfnamefont {R.~F.~L.}\ \bibnamefont {Vermeulen}}, \bibinfo {author} {\bibfnamefont {M.~J.}\ \bibnamefont {Tiggelman}}, \bibinfo {author} {\bibfnamefont {L.~d.~S.}\ \bibnamefont {Martins}}, \bibinfo {author} {\bibfnamefont {B.}~\bibnamefont {Dirkse}}, \bibinfo {author} {\bibfnamefont {S.}~\bibnamefont {Wehner}},\ and\ \bibinfo {author} {\bibfnamefont {R.}~\bibnamefont {Hanson}},\ }\bibfield  {title} {{\selectlanguage {en}\bibinfo {title} {Realization of a multinode quantum network of remote solid-state qubits}},\ }\href {https://doi.org/10.1126/science.abg1919} {\bibfield
  {journal} {\bibinfo  {journal} {Science}\ }\textbf {\bibinfo {volume} {372}},\ \bibinfo {pages} {6539} (\bibinfo {year} {2021})}\BibitemShut {NoStop}%
\bibitem [{\citenamefont {Hermans}\ \emph {et~al.}(2022)\citenamefont {Hermans}, \citenamefont {Pompili}, \citenamefont {Beukers}, \citenamefont {Baier}, \citenamefont {Borregaard},\ and\ \citenamefont {Hanson}}]{hermans_qubit_2022}%
  \BibitemOpen
  \bibfield  {author} {\bibinfo {author} {\bibfnamefont {S.~L.~N.}\ \bibnamefont {Hermans}}, \bibinfo {author} {\bibfnamefont {M.}~\bibnamefont {Pompili}}, \bibinfo {author} {\bibfnamefont {H.~K.~C.}\ \bibnamefont {Beukers}}, \bibinfo {author} {\bibfnamefont {S.}~\bibnamefont {Baier}}, \bibinfo {author} {\bibfnamefont {J.}~\bibnamefont {Borregaard}},\ and\ \bibinfo {author} {\bibfnamefont {R.}~\bibnamefont {Hanson}},\ }\bibfield  {title} {\bibinfo {title} {Qubit teleportation between non-neighbouring nodes in a quantum network},\ }\href {https://doi.org/10.1038/s41586-022-04697-y} {\bibfield  {journal} {\bibinfo  {journal} {Nature}\ }\textbf {\bibinfo {volume} {605}},\ \bibinfo {pages} {663} (\bibinfo {year} {2022})}\BibitemShut {NoStop}%
\bibitem [{\citenamefont {Li}\ \emph {et~al.}(2024)\citenamefont {Li}, \citenamefont {De~Santis}, \citenamefont {Harris}, \citenamefont {Chen}, \citenamefont {Gao}, \citenamefont {Christen}, \citenamefont {Choi}, \citenamefont {Trusheim}, \citenamefont {Song}, \citenamefont {Errando-Herranz}, \citenamefont {Du}, \citenamefont {Hu}, \citenamefont {Clark}, \citenamefont {Ibrahim}, \citenamefont {Gilbert}, \citenamefont {Han},\ and\ \citenamefont {Englund}}]{li_heterogeneous_2024}%
  \BibitemOpen
  \bibfield  {author} {\bibinfo {author} {\bibfnamefont {L.}~\bibnamefont {Li}}, \bibinfo {author} {\bibfnamefont {L.}~\bibnamefont {De~Santis}}, \bibinfo {author} {\bibfnamefont {I.~B.~W.}\ \bibnamefont {Harris}}, \bibinfo {author} {\bibfnamefont {K.~C.}\ \bibnamefont {Chen}}, \bibinfo {author} {\bibfnamefont {Y.}~\bibnamefont {Gao}}, \bibinfo {author} {\bibfnamefont {I.}~\bibnamefont {Christen}}, \bibinfo {author} {\bibfnamefont {H.}~\bibnamefont {Choi}}, \bibinfo {author} {\bibfnamefont {M.}~\bibnamefont {Trusheim}}, \bibinfo {author} {\bibfnamefont {Y.}~\bibnamefont {Song}}, \bibinfo {author} {\bibfnamefont {C.}~\bibnamefont {Errando-Herranz}}, \bibinfo {author} {\bibfnamefont {J.}~\bibnamefont {Du}}, \bibinfo {author} {\bibfnamefont {Y.}~\bibnamefont {Hu}}, \bibinfo {author} {\bibfnamefont {G.}~\bibnamefont {Clark}}, \bibinfo {author} {\bibfnamefont {M.~I.}\ \bibnamefont {Ibrahim}}, \bibinfo {author} {\bibfnamefont {G.}~\bibnamefont {Gilbert}}, \bibinfo {author} {\bibfnamefont {R.}~\bibnamefont {Han}},\
  and\ \bibinfo {author} {\bibfnamefont {D.}~\bibnamefont {Englund}},\ }\bibfield  {title} {\bibinfo {title} {Heterogeneous integration of spin–photon interfaces with a {CMOS} platform},\ }\href {https://doi.org/10.1038/s41586-024-07371-7} {\bibfield  {journal} {\bibinfo  {journal} {Nature}\ }\textbf {\bibinfo {volume} {630}},\ \bibinfo {pages} {70} (\bibinfo {year} {2024})}\BibitemShut {NoStop}%
\bibitem [{\citenamefont {Stolk}\ \emph {et~al.}(2024)\citenamefont {Stolk}, \citenamefont {Enden}, \citenamefont {Slater}, \citenamefont {Raa-Derckx}, \citenamefont {Botma}, \citenamefont {Rantwijk}, \citenamefont {Biemond}, \citenamefont {Hagen}, \citenamefont {Herfst}, \citenamefont {Koek}, \citenamefont {Meskers}, \citenamefont {Vollmer}, \citenamefont {Zwet}, \citenamefont {Markham}, \citenamefont {Edmonds}, \citenamefont {Geus}, \citenamefont {Elsen}, \citenamefont {Jungbluth}, \citenamefont {Haefner}, \citenamefont {Tresp}, \citenamefont {Stuhler}, \citenamefont {Ritter},\ and\ \citenamefont {Hanson}}]{stolk_metropolitan-scale_2024}%
  \BibitemOpen
  \bibfield  {author} {\bibinfo {author} {\bibfnamefont {A.~J.}\ \bibnamefont {Stolk}}, \bibinfo {author} {\bibfnamefont {K.~L. v.~d.}\ \bibnamefont {Enden}}, \bibinfo {author} {\bibfnamefont {M.-C.}\ \bibnamefont {Slater}}, \bibinfo {author} {\bibfnamefont {I.~t.}\ \bibnamefont {Raa-Derckx}}, \bibinfo {author} {\bibfnamefont {P.}~\bibnamefont {Botma}}, \bibinfo {author} {\bibfnamefont {J.~v.}\ \bibnamefont {Rantwijk}}, \bibinfo {author} {\bibfnamefont {J.~J.~B.}\ \bibnamefont {Biemond}}, \bibinfo {author} {\bibfnamefont {R.~A.~J.}\ \bibnamefont {Hagen}}, \bibinfo {author} {\bibfnamefont {R.~W.}\ \bibnamefont {Herfst}}, \bibinfo {author} {\bibfnamefont {W.~D.}\ \bibnamefont {Koek}}, \bibinfo {author} {\bibfnamefont {A.~J.~H.}\ \bibnamefont {Meskers}}, \bibinfo {author} {\bibfnamefont {R.}~\bibnamefont {Vollmer}}, \bibinfo {author} {\bibfnamefont {E.~J.~v.}\ \bibnamefont {Zwet}}, \bibinfo {author} {\bibfnamefont {M.}~\bibnamefont {Markham}}, \bibinfo {author} {\bibfnamefont {A.~M.}\ \bibnamefont {Edmonds}},
  \bibinfo {author} {\bibfnamefont {J.~F.}\ \bibnamefont {Geus}}, \bibinfo {author} {\bibfnamefont {F.}~\bibnamefont {Elsen}}, \bibinfo {author} {\bibfnamefont {B.}~\bibnamefont {Jungbluth}}, \bibinfo {author} {\bibfnamefont {C.}~\bibnamefont {Haefner}}, \bibinfo {author} {\bibfnamefont {C.}~\bibnamefont {Tresp}}, \bibinfo {author} {\bibfnamefont {J.}~\bibnamefont {Stuhler}}, \bibinfo {author} {\bibfnamefont {S.}~\bibnamefont {Ritter}},\ and\ \bibinfo {author} {\bibfnamefont {R.}~\bibnamefont {Hanson}},\ }\bibfield  {title} {\bibinfo {title} {Metropolitan-scale heralded entanglement of solid-state qubits},\ }\href {https://www.science.org/doi/10.1126/sciadv.adp6442} {\bibfield  {journal} {\bibinfo  {journal} {Science Advances}\ }\textbf {\bibinfo {volume} {10}},\ \bibinfo {pages} {44} (\bibinfo {year} {2024})}\BibitemShut {NoStop}%
\bibitem [{\citenamefont {Knaut}\ \emph {et~al.}(2024)\citenamefont {Knaut}, \citenamefont {Suleymanzade}, \citenamefont {Wei}, \citenamefont {Assumpcao}, \citenamefont {Stas}, \citenamefont {Huan}, \citenamefont {Machielse}, \citenamefont {Knall}, \citenamefont {Sutula}, \citenamefont {Baranes}, \citenamefont {Sinclair}, \citenamefont {De-Eknamkul}, \citenamefont {Levonian}, \citenamefont {Bhaskar}, \citenamefont {Park}, \citenamefont {Lončar},\ and\ \citenamefont {Lukin}}]{knaut_entanglement_2024}%
  \BibitemOpen
  \bibfield  {author} {\bibinfo {author} {\bibfnamefont {C.~M.}\ \bibnamefont {Knaut}}, \bibinfo {author} {\bibfnamefont {A.}~\bibnamefont {Suleymanzade}}, \bibinfo {author} {\bibfnamefont {Y.-C.}\ \bibnamefont {Wei}}, \bibinfo {author} {\bibfnamefont {D.~R.}\ \bibnamefont {Assumpcao}}, \bibinfo {author} {\bibfnamefont {P.-J.}\ \bibnamefont {Stas}}, \bibinfo {author} {\bibfnamefont {Y.~Q.}\ \bibnamefont {Huan}}, \bibinfo {author} {\bibfnamefont {B.}~\bibnamefont {Machielse}}, \bibinfo {author} {\bibfnamefont {E.~N.}\ \bibnamefont {Knall}}, \bibinfo {author} {\bibfnamefont {M.}~\bibnamefont {Sutula}}, \bibinfo {author} {\bibfnamefont {G.}~\bibnamefont {Baranes}}, \bibinfo {author} {\bibfnamefont {N.}~\bibnamefont {Sinclair}}, \bibinfo {author} {\bibfnamefont {C.}~\bibnamefont {De-Eknamkul}}, \bibinfo {author} {\bibfnamefont {D.~S.}\ \bibnamefont {Levonian}}, \bibinfo {author} {\bibfnamefont {M.~K.}\ \bibnamefont {Bhaskar}}, \bibinfo {author} {\bibfnamefont {H.}~\bibnamefont {Park}}, \bibinfo {author}
  {\bibfnamefont {M.}~\bibnamefont {Lončar}},\ and\ \bibinfo {author} {\bibfnamefont {M.~D.}\ \bibnamefont {Lukin}},\ }\bibfield  {title} {\bibinfo {title} {Entanglement of nanophotonic quantum memory nodes in a telecom network},\ }\href {https://doi.org/10.1038/s41586-024-07252-z} {\bibfield  {journal} {\bibinfo  {journal} {Nature}\ }\textbf {\bibinfo {volume} {629}},\ \bibinfo {pages} {573} (\bibinfo {year} {2024})}\BibitemShut {NoStop}%
\bibitem [{\citenamefont {Beukers}\ \emph {et~al.}(2024)\citenamefont {Beukers}, \citenamefont {Pasini}, \citenamefont {Choi}, \citenamefont {Englund}, \citenamefont {Hanson},\ and\ \citenamefont {Borregaard}}]{beukers_remote-entanglement_2024}%
  \BibitemOpen
  \bibfield  {author} {\bibinfo {author} {\bibfnamefont {H.~K.}\ \bibnamefont {Beukers}}, \bibinfo {author} {\bibfnamefont {M.}~\bibnamefont {Pasini}}, \bibinfo {author} {\bibfnamefont {H.}~\bibnamefont {Choi}}, \bibinfo {author} {\bibfnamefont {D.}~\bibnamefont {Englund}}, \bibinfo {author} {\bibfnamefont {R.}~\bibnamefont {Hanson}},\ and\ \bibinfo {author} {\bibfnamefont {J.}~\bibnamefont {Borregaard}},\ }\bibfield  {title} {{\selectlanguage {en}\bibinfo {title} {Remote-{Entanglement} {Protocols} for {Stationary} {Qubits} with {Photonic} {Interfaces}}},\ }\href {https://doi.org/10.1103/PRXQuantum.5.010202} {\bibfield  {journal} {\bibinfo  {journal} {PRX Quantum}\ }\textbf {\bibinfo {volume} {5}},\ \bibinfo {pages} {010202} (\bibinfo {year} {2024})}\BibitemShut {NoStop}%
\bibitem [{\citenamefont {Daiss}\ \emph {et~al.}(2021)\citenamefont {Daiss}, \citenamefont {Langenfeld}, \citenamefont {Welte}, \citenamefont {Distante}, \citenamefont {Thomas}, \citenamefont {Hartung}, \citenamefont {Morin},\ and\ \citenamefont {Rempe}}]{daiss_quantum-logic_2021}%
  \BibitemOpen
  \bibfield  {author} {\bibinfo {author} {\bibfnamefont {S.}~\bibnamefont {Daiss}}, \bibinfo {author} {\bibfnamefont {S.}~\bibnamefont {Langenfeld}}, \bibinfo {author} {\bibfnamefont {S.}~\bibnamefont {Welte}}, \bibinfo {author} {\bibfnamefont {E.}~\bibnamefont {Distante}}, \bibinfo {author} {\bibfnamefont {P.}~\bibnamefont {Thomas}}, \bibinfo {author} {\bibfnamefont {L.}~\bibnamefont {Hartung}}, \bibinfo {author} {\bibfnamefont {O.}~\bibnamefont {Morin}},\ and\ \bibinfo {author} {\bibfnamefont {G.}~\bibnamefont {Rempe}},\ }\bibfield  {title} {\bibinfo {title} {A {Quantum}-{Logic} {Gate} between {Distant} {Quantum}-{Network} {Modules}},\ }\href {https://doi.org/10.1126/science.abe3150} {\bibfield  {journal} {\bibinfo  {journal} {Science}\ }\textbf {\bibinfo {volume} {371}},\ \bibinfo {pages} {6529} (\bibinfo {year} {2021})}\BibitemShut {NoStop}%
\bibitem [{\citenamefont {Krutyanskiy}\ \emph {et~al.}(2019)\citenamefont {Krutyanskiy}, \citenamefont {Meraner}, \citenamefont {Schupp}, \citenamefont {Krcmarsky}, \citenamefont {Hainzer},\ and\ \citenamefont {Lanyon}}]{krutyanskiy_light-matter_2019}%
  \BibitemOpen
  \bibfield  {author} {\bibinfo {author} {\bibfnamefont {V.}~\bibnamefont {Krutyanskiy}}, \bibinfo {author} {\bibfnamefont {M.}~\bibnamefont {Meraner}}, \bibinfo {author} {\bibfnamefont {J.}~\bibnamefont {Schupp}}, \bibinfo {author} {\bibfnamefont {V.}~\bibnamefont {Krcmarsky}}, \bibinfo {author} {\bibfnamefont {H.}~\bibnamefont {Hainzer}},\ and\ \bibinfo {author} {\bibfnamefont {B.~P.}\ \bibnamefont {Lanyon}},\ }\bibfield  {title} {{\selectlanguage {en}\bibinfo {title} {Light-matter entanglement over 50 km of optical fibre}},\ }\href {https://doi.org/10.1038/s41534-019-0186-3} {\bibfield  {journal} {\bibinfo  {journal} {npj Quantum Inf}\ }\textbf {\bibinfo {volume} {5}},\ \bibinfo {pages} {72} (\bibinfo {year} {2019})}\BibitemShut {NoStop}%
\bibitem [{\citenamefont {Tchebotareva}\ \emph {et~al.}(2019)\citenamefont {Tchebotareva}, \citenamefont {Hermans}, \citenamefont {Humphreys}, \citenamefont {Voigt}, \citenamefont {Harmsma}, \citenamefont {Cheng}, \citenamefont {Verlaan}, \citenamefont {Dijkhuizen}, \citenamefont {De~Jong}, \citenamefont {Dréau},\ and\ \citenamefont {Hanson}}]{tchebotareva_entanglement_2019}%
  \BibitemOpen
  \bibfield  {author} {\bibinfo {author} {\bibfnamefont {A.}~\bibnamefont {Tchebotareva}}, \bibinfo {author} {\bibfnamefont {S.~L.}\ \bibnamefont {Hermans}}, \bibinfo {author} {\bibfnamefont {P.~C.}\ \bibnamefont {Humphreys}}, \bibinfo {author} {\bibfnamefont {D.}~\bibnamefont {Voigt}}, \bibinfo {author} {\bibfnamefont {P.~J.}\ \bibnamefont {Harmsma}}, \bibinfo {author} {\bibfnamefont {L.~K.}\ \bibnamefont {Cheng}}, \bibinfo {author} {\bibfnamefont {A.~L.}\ \bibnamefont {Verlaan}}, \bibinfo {author} {\bibfnamefont {N.}~\bibnamefont {Dijkhuizen}}, \bibinfo {author} {\bibfnamefont {W.}~\bibnamefont {De~Jong}}, \bibinfo {author} {\bibfnamefont {A.}~\bibnamefont {Dréau}},\ and\ \bibinfo {author} {\bibfnamefont {R.}~\bibnamefont {Hanson}},\ }\bibfield  {title} {{\selectlanguage {english}\bibinfo {title} {Entanglement between a {Diamond} {Spin} {Qubit} and a {Photonic} {Time}-{Bin} {Qubit} at {Telecom} {Wavelength}}},\ }\href {https://doi.org/10.1103/PhysRevLett.123.063601} {\bibfield  {journal} {\bibinfo
  {journal} {Phys. Rev. Lett.}\ }\textbf {\bibinfo {volume} {123}},\ \bibinfo {pages} {063601} (\bibinfo {year} {2019})}\BibitemShut {NoStop}%
\bibitem [{\citenamefont {Stolk}\ \emph {et~al.}(2022)\citenamefont {Stolk}, \citenamefont {van~der Enden}, \citenamefont {Roehsner}, \citenamefont {Teepe}, \citenamefont {Faes}, \citenamefont {Bradley}, \citenamefont {Cadot}, \citenamefont {van Rantwijk}, \citenamefont {te~Raa}, \citenamefont {Hagen}, \citenamefont {Verlaan}, \citenamefont {Biemond}, \citenamefont {Khorev}, \citenamefont {Vollmer}, \citenamefont {Markham}, \citenamefont {Edmonds}, \citenamefont {Morits}, \citenamefont {Taminiau}, \citenamefont {van Zwet},\ and\ \citenamefont {Hanson}}]{stolk_telecom-band_2022}%
  \BibitemOpen
  \bibfield  {author} {\bibinfo {author} {\bibfnamefont {A.}~\bibnamefont {Stolk}}, \bibinfo {author} {\bibfnamefont {K.}~\bibnamefont {van~der Enden}}, \bibinfo {author} {\bibfnamefont {M.-C.}\ \bibnamefont {Roehsner}}, \bibinfo {author} {\bibfnamefont {A.}~\bibnamefont {Teepe}}, \bibinfo {author} {\bibfnamefont {S.}~\bibnamefont {Faes}}, \bibinfo {author} {\bibfnamefont {C.}~\bibnamefont {Bradley}}, \bibinfo {author} {\bibfnamefont {S.}~\bibnamefont {Cadot}}, \bibinfo {author} {\bibfnamefont {J.}~\bibnamefont {van Rantwijk}}, \bibinfo {author} {\bibfnamefont {I.}~\bibnamefont {te~Raa}}, \bibinfo {author} {\bibfnamefont {R.}~\bibnamefont {Hagen}}, \bibinfo {author} {\bibfnamefont {A.}~\bibnamefont {Verlaan}}, \bibinfo {author} {\bibfnamefont {J.}~\bibnamefont {Biemond}}, \bibinfo {author} {\bibfnamefont {A.}~\bibnamefont {Khorev}}, \bibinfo {author} {\bibfnamefont {R.}~\bibnamefont {Vollmer}}, \bibinfo {author} {\bibfnamefont {M.}~\bibnamefont {Markham}}, \bibinfo {author} {\bibfnamefont {A.}~\bibnamefont
  {Edmonds}}, \bibinfo {author} {\bibfnamefont {J.}~\bibnamefont {Morits}}, \bibinfo {author} {\bibfnamefont {T.}~\bibnamefont {Taminiau}}, \bibinfo {author} {\bibfnamefont {E.}~\bibnamefont {van Zwet}},\ and\ \bibinfo {author} {\bibfnamefont {R.}~\bibnamefont {Hanson}},\ }\bibfield  {title} {\bibinfo {title} {Telecom-{Band} {Quantum} {Interference} of {Frequency}-{Converted} {Photons} from {Remote} {Detuned} {NV} {Centers}},\ }\href {https://doi.org/10.1103/PRXQuantum.3.020359} {\bibfield  {journal} {\bibinfo  {journal} {PRX Quantum}\ }\textbf {\bibinfo {volume} {3}},\ \bibinfo {pages} {020359} (\bibinfo {year} {2022})}\BibitemShut {NoStop}%
\bibitem [{\citenamefont {Iwasaki}\ \emph {et~al.}(2017)\citenamefont {Iwasaki}, \citenamefont {Miyamoto}, \citenamefont {Taniguchi}, \citenamefont {Siyushev}, \citenamefont {Metsch}, \citenamefont {Jelezko},\ and\ \citenamefont {Hatano}}]{iwasaki_tin-vacancy_2017}%
  \BibitemOpen
  \bibfield  {author} {\bibinfo {author} {\bibfnamefont {T.}~\bibnamefont {Iwasaki}}, \bibinfo {author} {\bibfnamefont {Y.}~\bibnamefont {Miyamoto}}, \bibinfo {author} {\bibfnamefont {T.}~\bibnamefont {Taniguchi}}, \bibinfo {author} {\bibfnamefont {P.}~\bibnamefont {Siyushev}}, \bibinfo {author} {\bibfnamefont {M.~H.}\ \bibnamefont {Metsch}}, \bibinfo {author} {\bibfnamefont {F.}~\bibnamefont {Jelezko}},\ and\ \bibinfo {author} {\bibfnamefont {M.}~\bibnamefont {Hatano}},\ }\bibfield  {title} {\bibinfo {title} {Tin-{Vacancy} {Quantum} {Emitters} in {Diamond}},\ }\href {https://doi.org/10.1103/PhysRevLett.119.253601} {\bibfield  {journal} {\bibinfo  {journal} {Phys. Rev. Lett.}\ }\textbf {\bibinfo {volume} {119}},\ \bibinfo {pages} {253601} (\bibinfo {year} {2017})}\BibitemShut {NoStop}%
\bibitem [{\citenamefont {Herrmann}\ \emph {et~al.}(2024)\citenamefont {Herrmann}, \citenamefont {Fischer}, \citenamefont {Brevoord}, \citenamefont {Sauerzapf}, \citenamefont {Wienhoven}, \citenamefont {Feije}, \citenamefont {Pasini}, \citenamefont {Eschen}, \citenamefont {Ruf}, \citenamefont {Weaver},\ and\ \citenamefont {Hanson}}]{herrmann_coherent_2024}%
  \BibitemOpen
  \bibfield  {author} {\bibinfo {author} {\bibfnamefont {Y.}~\bibnamefont {Herrmann}}, \bibinfo {author} {\bibfnamefont {J.}~\bibnamefont {Fischer}}, \bibinfo {author} {\bibfnamefont {J.~M.}\ \bibnamefont {Brevoord}}, \bibinfo {author} {\bibfnamefont {C.}~\bibnamefont {Sauerzapf}}, \bibinfo {author} {\bibfnamefont {L.~G.}\ \bibnamefont {Wienhoven}}, \bibinfo {author} {\bibfnamefont {L.~J.}\ \bibnamefont {Feije}}, \bibinfo {author} {\bibfnamefont {M.}~\bibnamefont {Pasini}}, \bibinfo {author} {\bibfnamefont {M.}~\bibnamefont {Eschen}}, \bibinfo {author} {\bibfnamefont {M.}~\bibnamefont {Ruf}}, \bibinfo {author} {\bibfnamefont {M.~J.}\ \bibnamefont {Weaver}},\ and\ \bibinfo {author} {\bibfnamefont {R.}~\bibnamefont {Hanson}},\ }\bibfield  {title} {\bibinfo {title} {Coherent {Coupling} of a {Diamond} {Tin}-{Vacancy} {Center} to a {Tunable} {Open} {Microcavity}},\ }\href {https://doi.org/10.1103/PhysRevX.14.041013} {\bibfield  {journal} {\bibinfo  {journal} {Phys. Rev. X}\ }\textbf {\bibinfo {volume} {14}},\
  \bibinfo {pages} {041013} (\bibinfo {year} {2024})}\BibitemShut {NoStop}%
\bibitem [{\citenamefont {Karapatzakis}\ \emph {et~al.}(2024)\citenamefont {Karapatzakis}, \citenamefont {Resch}, \citenamefont {Schrodin}, \citenamefont {Fuchs}, \citenamefont {Kieschnick}, \citenamefont {Heupel}, \citenamefont {Kussi}, \citenamefont {Sürgers}, \citenamefont {Popov}, \citenamefont {Meijer}, \citenamefont {Becher}, \citenamefont {Wernsdorfer},\ and\ \citenamefont {Hunger}}]{karapatzakis_microwave_2024}%
  \BibitemOpen
  \bibfield  {author} {\bibinfo {author} {\bibfnamefont {I.}~\bibnamefont {Karapatzakis}}, \bibinfo {author} {\bibfnamefont {J.}~\bibnamefont {Resch}}, \bibinfo {author} {\bibfnamefont {M.}~\bibnamefont {Schrodin}}, \bibinfo {author} {\bibfnamefont {P.}~\bibnamefont {Fuchs}}, \bibinfo {author} {\bibfnamefont {M.}~\bibnamefont {Kieschnick}}, \bibinfo {author} {\bibfnamefont {J.}~\bibnamefont {Heupel}}, \bibinfo {author} {\bibfnamefont {L.}~\bibnamefont {Kussi}}, \bibinfo {author} {\bibfnamefont {C.}~\bibnamefont {Sürgers}}, \bibinfo {author} {\bibfnamefont {C.}~\bibnamefont {Popov}}, \bibinfo {author} {\bibfnamefont {J.}~\bibnamefont {Meijer}}, \bibinfo {author} {\bibfnamefont {C.}~\bibnamefont {Becher}}, \bibinfo {author} {\bibfnamefont {W.}~\bibnamefont {Wernsdorfer}},\ and\ \bibinfo {author} {\bibfnamefont {D.}~\bibnamefont {Hunger}},\ }\bibfield  {title} {\bibinfo {title} {Microwave {Control} of the {Tin}-{Vacancy} {Spin} {Qubit} in {Diamond} with a {Superconducting} {Waveguide}},\ }\href
  {https://doi.org/10.1103/PhysRevX.14.031036} {\bibfield  {journal} {\bibinfo  {journal} {Phys. Rev. X}\ }\textbf {\bibinfo {volume} {14}},\ \bibinfo {pages} {031036} (\bibinfo {year} {2024})}\BibitemShut {NoStop}%
\bibitem [{\citenamefont {Guo}\ \emph {et~al.}(2023)\citenamefont {Guo}, \citenamefont {Stramma}, \citenamefont {Li}, \citenamefont {Roth}, \citenamefont {Huang}, \citenamefont {Jin}, \citenamefont {Parker}, \citenamefont {Arjona~Martínez}, \citenamefont {Shofer}, \citenamefont {Michaels}, \citenamefont {Purser}, \citenamefont {Appel}, \citenamefont {Alexeev}, \citenamefont {Liu}, \citenamefont {Ferrari}, \citenamefont {Awschalom}, \citenamefont {Delegan}, \citenamefont {Pingault}, \citenamefont {Galli}, \citenamefont {Heremans}, \citenamefont {Atatüre},\ and\ \citenamefont {High}}]{guo_microwave-based_2023}%
  \BibitemOpen
  \bibfield  {author} {\bibinfo {author} {\bibfnamefont {X.}~\bibnamefont {Guo}}, \bibinfo {author} {\bibfnamefont {A.~M.}\ \bibnamefont {Stramma}}, \bibinfo {author} {\bibfnamefont {Z.}~\bibnamefont {Li}}, \bibinfo {author} {\bibfnamefont {W.~G.}\ \bibnamefont {Roth}}, \bibinfo {author} {\bibfnamefont {B.}~\bibnamefont {Huang}}, \bibinfo {author} {\bibfnamefont {Y.}~\bibnamefont {Jin}}, \bibinfo {author} {\bibfnamefont {R.~A.}\ \bibnamefont {Parker}}, \bibinfo {author} {\bibfnamefont {J.}~\bibnamefont {Arjona~Martínez}}, \bibinfo {author} {\bibfnamefont {N.}~\bibnamefont {Shofer}}, \bibinfo {author} {\bibfnamefont {C.~P.}\ \bibnamefont {Michaels}}, \bibinfo {author} {\bibfnamefont {C.~P.}\ \bibnamefont {Purser}}, \bibinfo {author} {\bibfnamefont {M.~H.}\ \bibnamefont {Appel}}, \bibinfo {author} {\bibfnamefont {E.~M.}\ \bibnamefont {Alexeev}}, \bibinfo {author} {\bibfnamefont {T.}~\bibnamefont {Liu}}, \bibinfo {author} {\bibfnamefont {A.~C.}\ \bibnamefont {Ferrari}}, \bibinfo {author} {\bibfnamefont {D.~D.}\
  \bibnamefont {Awschalom}}, \bibinfo {author} {\bibfnamefont {N.}~\bibnamefont {Delegan}}, \bibinfo {author} {\bibfnamefont {B.}~\bibnamefont {Pingault}}, \bibinfo {author} {\bibfnamefont {G.}~\bibnamefont {Galli}}, \bibinfo {author} {\bibfnamefont {F.~J.}\ \bibnamefont {Heremans}}, \bibinfo {author} {\bibfnamefont {M.}~\bibnamefont {Atatüre}},\ and\ \bibinfo {author} {\bibfnamefont {A.~A.}\ \bibnamefont {High}},\ }\bibfield  {title} {\bibinfo {title} {Microwave-{Based} {Quantum} {Control} and {Coherence} {Protection} of {Tin}-{Vacancy} {Spin} {Qubits} in a {Strain}-{Tuned} {Diamond}-{Membrane} {Heterostructure}},\ }\href {https://doi.org/10.1103/PhysRevX.13.041037} {\bibfield  {journal} {\bibinfo  {journal} {Phys. Rev. X}\ }\textbf {\bibinfo {volume} {13}},\ \bibinfo {pages} {041037} (\bibinfo {year} {2023})}\BibitemShut {NoStop}%
\bibitem [{\citenamefont {Rosenthal}\ \emph {et~al.}(2023)\citenamefont {Rosenthal}, \citenamefont {Anderson}, \citenamefont {Kleidermacher}, \citenamefont {Stein}, \citenamefont {Lee}, \citenamefont {Grzesik}, \citenamefont {Scuri}, \citenamefont {Rugar}, \citenamefont {Riedel}, \citenamefont {Aghaeimeibodi}, \citenamefont {Ahn}, \citenamefont {Van~Gasse},\ and\ \citenamefont {Vučković}}]{rosenthal_microwave_2023}%
  \BibitemOpen
  \bibfield  {author} {\bibinfo {author} {\bibfnamefont {E.~I.}\ \bibnamefont {Rosenthal}}, \bibinfo {author} {\bibfnamefont {C.~P.}\ \bibnamefont {Anderson}}, \bibinfo {author} {\bibfnamefont {H.~C.}\ \bibnamefont {Kleidermacher}}, \bibinfo {author} {\bibfnamefont {A.~J.}\ \bibnamefont {Stein}}, \bibinfo {author} {\bibfnamefont {H.}~\bibnamefont {Lee}}, \bibinfo {author} {\bibfnamefont {J.}~\bibnamefont {Grzesik}}, \bibinfo {author} {\bibfnamefont {G.}~\bibnamefont {Scuri}}, \bibinfo {author} {\bibfnamefont {A.~E.}\ \bibnamefont {Rugar}}, \bibinfo {author} {\bibfnamefont {D.}~\bibnamefont {Riedel}}, \bibinfo {author} {\bibfnamefont {S.}~\bibnamefont {Aghaeimeibodi}}, \bibinfo {author} {\bibfnamefont {G.~H.}\ \bibnamefont {Ahn}}, \bibinfo {author} {\bibfnamefont {K.}~\bibnamefont {Van~Gasse}},\ and\ \bibinfo {author} {\bibfnamefont {J.}~\bibnamefont {Vučković}},\ }\bibfield  {title} {\bibinfo {title} {Microwave {Spin} {Control} of a {Tin}-{Vacancy} {Qubit} in {Diamond}},\ }\href
  {https://doi.org/10.1103/PhysRevX.13.031022} {\bibfield  {journal} {\bibinfo  {journal} {Phys. Rev. X}\ }\textbf {\bibinfo {volume} {13}},\ \bibinfo {pages} {031022} (\bibinfo {year} {2023})}\BibitemShut {NoStop}%
\bibitem [{\citenamefont {Thiering}\ and\ \citenamefont {Gali}(2018)}]{thiering_ab_2018}%
  \BibitemOpen
  \bibfield  {author} {\bibinfo {author} {\bibfnamefont {G.}~\bibnamefont {Thiering}}\ and\ \bibinfo {author} {\bibfnamefont {A.}~\bibnamefont {Gali}},\ }\bibfield  {title} {\bibinfo {title} {Ab {Initio} {Magneto}-{Optical} {Spectrum} of {Group}-{IV} {Vacancy} {Color} {Centers} in {Diamond}},\ }\href {https://doi.org/10.1103/PhysRevX.8.021063} {\bibfield  {journal} {\bibinfo  {journal} {Physical Review X}\ }\textbf {\bibinfo {volume} {8}},\ \bibinfo {pages} {021063} (\bibinfo {year} {2018})}\BibitemShut {NoStop}%
\bibitem [{\citenamefont {Rugar}\ \emph {et~al.}(2021)\citenamefont {Rugar}, \citenamefont {Aghaeimeibodi}, \citenamefont {Riedel}, \citenamefont {Dory}, \citenamefont {Lu}, \citenamefont {McQuade}, \citenamefont {Shen}, \citenamefont {Melosh},\ and\ \citenamefont {Vučković}}]{rugar_quantum_2021}%
  \BibitemOpen
  \bibfield  {author} {\bibinfo {author} {\bibfnamefont {A.~E.}\ \bibnamefont {Rugar}}, \bibinfo {author} {\bibfnamefont {S.}~\bibnamefont {Aghaeimeibodi}}, \bibinfo {author} {\bibfnamefont {D.}~\bibnamefont {Riedel}}, \bibinfo {author} {\bibfnamefont {C.}~\bibnamefont {Dory}}, \bibinfo {author} {\bibfnamefont {H.}~\bibnamefont {Lu}}, \bibinfo {author} {\bibfnamefont {P.~J.}\ \bibnamefont {McQuade}}, \bibinfo {author} {\bibfnamefont {Z.-X.}\ \bibnamefont {Shen}}, \bibinfo {author} {\bibfnamefont {N.~A.}\ \bibnamefont {Melosh}},\ and\ \bibinfo {author} {\bibfnamefont {J.}~\bibnamefont {Vučković}},\ }\bibfield  {title} {\bibinfo {title} {Quantum {Photonic} {Interface} for {Tin}-{Vacancy} {Centers} in {Diamond}},\ }\href {https://doi.org/10.1103/PhysRevX.11.031021} {\bibfield  {journal} {\bibinfo  {journal} {Phys. Rev. X}\ }\textbf {\bibinfo {volume} {11}},\ \bibinfo {pages} {031021} (\bibinfo {year} {2021})}\BibitemShut {NoStop}%
\bibitem [{\citenamefont {Arjona~Martínez}\ \emph {et~al.}(2022)\citenamefont {Arjona~Martínez}, \citenamefont {Parker}, \citenamefont {Chen}, \citenamefont {Purser}, \citenamefont {Li}, \citenamefont {Michaels}, \citenamefont {Stramma}, \citenamefont {Debroux}, \citenamefont {Harris}, \citenamefont {Hayhurst~Appel}, \citenamefont {Nichols}, \citenamefont {Trusheim}, \citenamefont {Gangloff}, \citenamefont {Englund},\ and\ \citenamefont {Atatüre}}]{arjona_martinez_photonic_2022}%
  \BibitemOpen
  \bibfield  {author} {\bibinfo {author} {\bibfnamefont {J.}~\bibnamefont {Arjona~Martínez}}, \bibinfo {author} {\bibfnamefont {R.~A.}\ \bibnamefont {Parker}}, \bibinfo {author} {\bibfnamefont {K.~C.}\ \bibnamefont {Chen}}, \bibinfo {author} {\bibfnamefont {C.~M.}\ \bibnamefont {Purser}}, \bibinfo {author} {\bibfnamefont {L.}~\bibnamefont {Li}}, \bibinfo {author} {\bibfnamefont {C.~P.}\ \bibnamefont {Michaels}}, \bibinfo {author} {\bibfnamefont {A.~M.}\ \bibnamefont {Stramma}}, \bibinfo {author} {\bibfnamefont {R.}~\bibnamefont {Debroux}}, \bibinfo {author} {\bibfnamefont {I.~B.}\ \bibnamefont {Harris}}, \bibinfo {author} {\bibfnamefont {M.}~\bibnamefont {Hayhurst~Appel}}, \bibinfo {author} {\bibfnamefont {E.~C.}\ \bibnamefont {Nichols}}, \bibinfo {author} {\bibfnamefont {M.~E.}\ \bibnamefont {Trusheim}}, \bibinfo {author} {\bibfnamefont {D.~A.}\ \bibnamefont {Gangloff}}, \bibinfo {author} {\bibfnamefont {D.}~\bibnamefont {Englund}},\ and\ \bibinfo {author} {\bibfnamefont {M.}~\bibnamefont {Atatüre}},\
  }\bibfield  {title} {\bibinfo {title} {Photonic {Indistinguishability} of the {Tin}-{Vacancy} {Center} in {Nanostructured} {Diamond}},\ }\href {https://doi.org/10.1103/PhysRevLett.129.173603} {\bibfield  {journal} {\bibinfo  {journal} {Phys. Rev. Lett.}\ }\textbf {\bibinfo {volume} {129}},\ \bibinfo {pages} {17} (\bibinfo {year} {2022})}\BibitemShut {NoStop}%
\bibitem [{\citenamefont {Pasini}\ \emph {et~al.}(2024)\citenamefont {Pasini}, \citenamefont {Codreanu}, \citenamefont {Turan}, \citenamefont {Riera~Moral}, \citenamefont {Primavera}, \citenamefont {De~Santis}, \citenamefont {Beukers}, \citenamefont {Brevoord}, \citenamefont {Waas}, \citenamefont {Borregaard},\ and\ \citenamefont {Hanson}}]{pasini_nonlinear_2024}%
  \BibitemOpen
  \bibfield  {author} {\bibinfo {author} {\bibfnamefont {M.}~\bibnamefont {Pasini}}, \bibinfo {author} {\bibfnamefont {N.}~\bibnamefont {Codreanu}}, \bibinfo {author} {\bibfnamefont {T.}~\bibnamefont {Turan}}, \bibinfo {author} {\bibfnamefont {A.}~\bibnamefont {Riera~Moral}}, \bibinfo {author} {\bibfnamefont {C.~F.}\ \bibnamefont {Primavera}}, \bibinfo {author} {\bibfnamefont {L.}~\bibnamefont {De~Santis}}, \bibinfo {author} {\bibfnamefont {H.~K.~C.}\ \bibnamefont {Beukers}}, \bibinfo {author} {\bibfnamefont {J.~M.}\ \bibnamefont {Brevoord}}, \bibinfo {author} {\bibfnamefont {C.}~\bibnamefont {Waas}}, \bibinfo {author} {\bibfnamefont {J.}~\bibnamefont {Borregaard}},\ and\ \bibinfo {author} {\bibfnamefont {R.}~\bibnamefont {Hanson}},\ }\bibfield  {title} {\bibinfo {title} {Nonlinear {Quantum} {Photonics} with a {Tin}-{Vacancy} {Center} {Coupled} to a {One}-{Dimensional} {Diamond} {Waveguide}},\ }\href {https://doi.org/10.1103/PhysRevLett.133.023603} {\bibfield  {journal} {\bibinfo  {journal} {Phys. Rev.
  Lett.}\ }\textbf {\bibinfo {volume} {133}},\ \bibinfo {pages} {023603} (\bibinfo {year} {2024})}\BibitemShut {NoStop}%
\bibitem [{\citenamefont {Clark}\ \emph {et~al.}(2024)\citenamefont {Clark}, \citenamefont {Raniwala}, \citenamefont {Koppa}, \citenamefont {Chen}, \citenamefont {Leenheer}, \citenamefont {Zimmermann}, \citenamefont {Dong}, \citenamefont {Li}, \citenamefont {Wen}, \citenamefont {Dominguez}, \citenamefont {Trusheim}, \citenamefont {Gilbert}, \citenamefont {Eichenfield},\ and\ \citenamefont {Englund}}]{clark_nanoelectromechanical_2024}%
  \BibitemOpen
  \bibfield  {author} {\bibinfo {author} {\bibfnamefont {G.}~\bibnamefont {Clark}}, \bibinfo {author} {\bibfnamefont {H.}~\bibnamefont {Raniwala}}, \bibinfo {author} {\bibfnamefont {M.}~\bibnamefont {Koppa}}, \bibinfo {author} {\bibfnamefont {K.}~\bibnamefont {Chen}}, \bibinfo {author} {\bibfnamefont {A.}~\bibnamefont {Leenheer}}, \bibinfo {author} {\bibfnamefont {M.}~\bibnamefont {Zimmermann}}, \bibinfo {author} {\bibfnamefont {M.}~\bibnamefont {Dong}}, \bibinfo {author} {\bibfnamefont {L.}~\bibnamefont {Li}}, \bibinfo {author} {\bibfnamefont {Y.~H.}\ \bibnamefont {Wen}}, \bibinfo {author} {\bibfnamefont {D.}~\bibnamefont {Dominguez}}, \bibinfo {author} {\bibfnamefont {M.}~\bibnamefont {Trusheim}}, \bibinfo {author} {\bibfnamefont {G.}~\bibnamefont {Gilbert}}, \bibinfo {author} {\bibfnamefont {M.}~\bibnamefont {Eichenfield}},\ and\ \bibinfo {author} {\bibfnamefont {D.}~\bibnamefont {Englund}},\ }\bibfield  {title} {\bibinfo {title} {Nanoelectromechanical {Control} of {Spin}–{Photon} {Interfaces} in a
  {Hybrid} {Quantum} {System} on {Chip}},\ }\href {https://doi.org/10.1021/acs.nanolett.3c04301} {\bibfield  {journal} {\bibinfo  {journal} {Nano Letters}\ }\textbf {\bibinfo {volume} {24}},\ \bibinfo {pages} {4} (\bibinfo {year} {2024})}\BibitemShut {NoStop}%
\bibitem [{\citenamefont {Geus}\ \emph {et~al.}(2024)\citenamefont {Geus}, \citenamefont {Elsen}, \citenamefont {Nyga}, \citenamefont {Stolk}, \citenamefont {{van der Enden}}, \citenamefont {{van Zwet}}, \citenamefont {Haefner}, \citenamefont {Hanson},\ and\ \citenamefont {Jungbluth}}]{Geus.2024}%
  \BibitemOpen
  \bibfield  {author} {\bibinfo {author} {\bibfnamefont {J.~F.}\ \bibnamefont {Geus}}, \bibinfo {author} {\bibfnamefont {F.}~\bibnamefont {Elsen}}, \bibinfo {author} {\bibfnamefont {S.}~\bibnamefont {Nyga}}, \bibinfo {author} {\bibfnamefont {A.~J.}\ \bibnamefont {Stolk}}, \bibinfo {author} {\bibfnamefont {K.~L.}\ \bibnamefont {{van der Enden}}}, \bibinfo {author} {\bibfnamefont {E.~J.}\ \bibnamefont {{van Zwet}}}, \bibinfo {author} {\bibfnamefont {C.}~\bibnamefont {Haefner}}, \bibinfo {author} {\bibfnamefont {R.}~\bibnamefont {Hanson}},\ and\ \bibinfo {author} {\bibfnamefont {B.}~\bibnamefont {Jungbluth}},\ }\bibfield  {title} {\bibinfo {title} {Low-noise short-wavelength pumped frequency downconversion for quantum frequency converters},\ }\href {https://doi.org/10.1364/OPTICAQ.515769} {\bibfield  {journal} {\bibinfo  {journal} {Optica Quantum}\ }\textbf {\bibinfo {volume} {2}},\ \bibinfo {pages} {189} (\bibinfo {year} {2024})}\BibitemShut {NoStop}%
\bibitem [{\citenamefont {H{\"a}nsch}\ and\ \citenamefont {Couillaud}(1980)}]{Hansch.1980}%
  \BibitemOpen
  \bibfield  {author} {\bibinfo {author} {\bibfnamefont {T.~W.}\ \bibnamefont {H{\"a}nsch}}\ and\ \bibinfo {author} {\bibfnamefont {B.}~\bibnamefont {Couillaud}},\ }\bibfield  {title} {\bibinfo {title} {Laser frequency stabilization by polarization spectroscopy of a reflecting reference cavity},\ }\href {https://doi.org/10.1016/0030-4018(80)90069-3} {\bibfield  {journal} {\bibinfo  {journal} {Optics Communications}\ }\textbf {\bibinfo {volume} {35}},\ \bibinfo {pages} {441} (\bibinfo {year} {1980})}\BibitemShut {NoStop}%
\bibitem [{\citenamefont {Roussev}\ \emph {et~al.}(2004)\citenamefont {Roussev}, \citenamefont {Langrock}, \citenamefont {Kurz},\ and\ \citenamefont {Fejer}}]{Roussev.2004}%
  \BibitemOpen
  \bibfield  {author} {\bibinfo {author} {\bibfnamefont {R.~V.}\ \bibnamefont {Roussev}}, \bibinfo {author} {\bibfnamefont {C.}~\bibnamefont {Langrock}}, \bibinfo {author} {\bibfnamefont {J.~R.}\ \bibnamefont {Kurz}},\ and\ \bibinfo {author} {\bibfnamefont {M.~M.}\ \bibnamefont {Fejer}},\ }\bibfield  {title} {\bibinfo {title} {Periodically poled lithium niobate waveguide sum-frequency generator for efficient single-photon detection at communication wavelengths},\ }\href {https://doi.org/10.1364/OL.29.001518} {\bibfield  {journal} {\bibinfo  {journal} {Optics letters}\ }\textbf {\bibinfo {volume} {29}},\ \bibinfo {pages} {1518} (\bibinfo {year} {2004})}\BibitemShut {NoStop}%
\bibitem [{\citenamefont {Brevoord}\ \emph {et~al.}(2025{\natexlab{a}})\citenamefont {Brevoord}, \citenamefont {Wienhoven}, \citenamefont {Codreanu}, \citenamefont {Ishiguro}, \citenamefont {van Leeuwen}, \citenamefont {Iuliano}, \citenamefont {De~Santis}, \citenamefont {Waas}, \citenamefont {Beukers}, \citenamefont {Turan}, \citenamefont {Errando-Herranz}, \citenamefont {Kawaguchi},\ and\ \citenamefont {Hanson}}]{brevoord_large-range_2025}%
  \BibitemOpen
  \bibfield  {author} {\bibinfo {author} {\bibfnamefont {J.~M.}\ \bibnamefont {Brevoord}}, \bibinfo {author} {\bibfnamefont {L.~G.~C.}\ \bibnamefont {Wienhoven}}, \bibinfo {author} {\bibfnamefont {N.}~\bibnamefont {Codreanu}}, \bibinfo {author} {\bibfnamefont {T.}~\bibnamefont {Ishiguro}}, \bibinfo {author} {\bibfnamefont {E.}~\bibnamefont {van Leeuwen}}, \bibinfo {author} {\bibfnamefont {M.}~\bibnamefont {Iuliano}}, \bibinfo {author} {\bibfnamefont {L.}~\bibnamefont {De~Santis}}, \bibinfo {author} {\bibfnamefont {C.}~\bibnamefont {Waas}}, \bibinfo {author} {\bibfnamefont {H.~K.~C.}\ \bibnamefont {Beukers}}, \bibinfo {author} {\bibfnamefont {T.}~\bibnamefont {Turan}}, \bibinfo {author} {\bibfnamefont {C.}~\bibnamefont {Errando-Herranz}}, \bibinfo {author} {\bibfnamefont {K.}~\bibnamefont {Kawaguchi}},\ and\ \bibinfo {author} {\bibfnamefont {R.}~\bibnamefont {Hanson}},\ }\bibfield  {title} {{\selectlanguage {en}\bibinfo {title} {Large-range tuning and stabilization of the optical transition of diamond
  tin-vacancy centers by in situ strain control}},\ }\href {https://doi.org/10.1063/5.0251211} {\bibfield  {journal} {\bibinfo  {journal} {Appl. Phys. Lett.}\ }\textbf {\bibinfo {volume} {126}},\ \bibinfo {pages} {17} (\bibinfo {year} {2025}{\natexlab{a}})}\BibitemShut {NoStop}%
\bibitem [{\citenamefont {Görlitz}\ \emph {et~al.}(2020)\citenamefont {Görlitz}, \citenamefont {Herrmann}, \citenamefont {Thiering}, \citenamefont {Fuchs}, \citenamefont {Gandil}, \citenamefont {Iwasaki}, \citenamefont {Taniguchi}, \citenamefont {Kieschnick}, \citenamefont {Meijer}, \citenamefont {Hatano}, \citenamefont {Gali},\ and\ \citenamefont {Becher}}]{gorlitz_spectroscopic_2020}%
  \BibitemOpen
  \bibfield  {author} {\bibinfo {author} {\bibfnamefont {J.}~\bibnamefont {Görlitz}}, \bibinfo {author} {\bibfnamefont {D.}~\bibnamefont {Herrmann}}, \bibinfo {author} {\bibfnamefont {G.}~\bibnamefont {Thiering}}, \bibinfo {author} {\bibfnamefont {P.}~\bibnamefont {Fuchs}}, \bibinfo {author} {\bibfnamefont {M.}~\bibnamefont {Gandil}}, \bibinfo {author} {\bibfnamefont {T.}~\bibnamefont {Iwasaki}}, \bibinfo {author} {\bibfnamefont {T.}~\bibnamefont {Taniguchi}}, \bibinfo {author} {\bibfnamefont {M.}~\bibnamefont {Kieschnick}}, \bibinfo {author} {\bibfnamefont {J.}~\bibnamefont {Meijer}}, \bibinfo {author} {\bibfnamefont {M.}~\bibnamefont {Hatano}}, \bibinfo {author} {\bibfnamefont {A.}~\bibnamefont {Gali}},\ and\ \bibinfo {author} {\bibfnamefont {C.}~\bibnamefont {Becher}},\ }\bibfield  {title} {\bibinfo {title} {Spectroscopic investigations of negatively charged tin-vacancy centres in diamond},\ }\href {https://doi.org/10.1088/1367-2630/ab6631} {\bibfield  {journal} {\bibinfo  {journal} {New J. Phys.}\
  }\textbf {\bibinfo {volume} {22}},\ \bibinfo {pages} {013048} (\bibinfo {year} {2020})}\BibitemShut {NoStop}%
\bibitem [{\citenamefont {Narita}\ \emph {et~al.}(2023)\citenamefont {Narita}, \citenamefont {Wang}, \citenamefont {Ikeda}, \citenamefont {Oba}, \citenamefont {Miyamoto}, \citenamefont {Taniguchi}, \citenamefont {Onoda}, \citenamefont {Hatano},\ and\ \citenamefont {Iwasaki}}]{narita_multiple_2023}%
  \BibitemOpen
  \bibfield  {author} {\bibinfo {author} {\bibfnamefont {Y.}~\bibnamefont {Narita}}, \bibinfo {author} {\bibfnamefont {P.}~\bibnamefont {Wang}}, \bibinfo {author} {\bibfnamefont {K.}~\bibnamefont {Ikeda}}, \bibinfo {author} {\bibfnamefont {K.}~\bibnamefont {Oba}}, \bibinfo {author} {\bibfnamefont {Y.}~\bibnamefont {Miyamoto}}, \bibinfo {author} {\bibfnamefont {T.}~\bibnamefont {Taniguchi}}, \bibinfo {author} {\bibfnamefont {S.}~\bibnamefont {Onoda}}, \bibinfo {author} {\bibfnamefont {M.}~\bibnamefont {Hatano}},\ and\ \bibinfo {author} {\bibfnamefont {T.}~\bibnamefont {Iwasaki}},\ }\bibfield  {title} {{\selectlanguage {en}\bibinfo {title} {Multiple {Tin}-{Vacancy} {Centers} in {Diamond} with {Nearly} {Identical} {Photon} {Frequency} and {Linewidth}}},\ }\href {https://doi.org/10.1103/PhysRevApplied.19.024061} {\bibfield  {journal} {\bibinfo  {journal} {Phys. Rev. Applied}\ }\textbf {\bibinfo {volume} {19}},\ \bibinfo {pages} {024061} (\bibinfo {year} {2023})}\BibitemShut {NoStop}%
\bibitem [{\citenamefont {Bushmakin}\ \emph {et~al.}(2024)\citenamefont {Bushmakin}, \citenamefont {Berg}, \citenamefont {Sauerzapf}, \citenamefont {Jayaram}, \citenamefont {Denisenko}, \citenamefont {Vorobyov}, \citenamefont {Gerhardt}, \citenamefont {Liu},\ and\ \citenamefont {Wrachtrup}}]{bushmakin_two-photon_2024}%
  \BibitemOpen
  \bibfield  {author} {\bibinfo {author} {\bibfnamefont {V.}~\bibnamefont {Bushmakin}}, \bibinfo {author} {\bibfnamefont {O.~v.}\ \bibnamefont {Berg}}, \bibinfo {author} {\bibfnamefont {C.}~\bibnamefont {Sauerzapf}}, \bibinfo {author} {\bibfnamefont {S.}~\bibnamefont {Jayaram}}, \bibinfo {author} {\bibfnamefont {A.}~\bibnamefont {Denisenko}}, \bibinfo {author} {\bibfnamefont {V.}~\bibnamefont {Vorobyov}}, \bibinfo {author} {\bibfnamefont {I.}~\bibnamefont {Gerhardt}}, \bibinfo {author} {\bibfnamefont {D.}~\bibnamefont {Liu}},\ and\ \bibinfo {author} {\bibfnamefont {J.}~\bibnamefont {Wrachtrup}},\ }\href {https://doi.org/10.48550/arXiv.2412.17539} {{\selectlanguage {en}\bibinfo {title} {Two-{Photon} {Interference} of {Photons} from {Remote} {Tin}-{Vacancy} {Centers} in {Diamond}}}} (\bibinfo {year} {2024}),\ \bibinfo {note} {arXiv:2412.17539}\BibitemShut {NoStop}%
\bibitem [{\citenamefont {Hong}\ and\ \citenamefont {Mandel}(1985)}]{Hong.1985}%
  \BibitemOpen
  \bibfield  {author} {\bibinfo {author} {\bibfnamefont {C.~K.}\ \bibnamefont {Hong}}\ and\ \bibinfo {author} {\bibfnamefont {L.}~\bibnamefont {Mandel}},\ }\bibfield  {title} {\bibinfo {title} {Theory of parametric frequency down conversion of light},\ }\href {https://doi.org/10.1103/PhysRevA.31.2409} {\bibfield  {journal} {\bibinfo  {journal} {Physical review. A, General physics}\ }\textbf {\bibinfo {volume} {31}},\ \bibinfo {pages} {2409} (\bibinfo {year} {1985})}\BibitemShut {NoStop}%
\bibitem [{\citenamefont {Ruf}\ \emph {et~al.}(2021)\citenamefont {Ruf}, \citenamefont {Weaver}, \citenamefont {van Dam},\ and\ \citenamefont {Hanson}}]{ruf_resonant_2021}%
  \BibitemOpen
  \bibfield  {author} {\bibinfo {author} {\bibfnamefont {M.}~\bibnamefont {Ruf}}, \bibinfo {author} {\bibfnamefont {M.}~\bibnamefont {Weaver}}, \bibinfo {author} {\bibfnamefont {S.}~\bibnamefont {van Dam}},\ and\ \bibinfo {author} {\bibfnamefont {R.}~\bibnamefont {Hanson}},\ }\bibfield  {title} {{\selectlanguage {english}\bibinfo {title} {Resonant {Excitation} and {Purcell} {Enhancement} of {Coherent} {Nitrogen}-{Vacancy} {Centers} {Coupled} to a {Fabry}-{Perot} {Microcavity}}},\ }\href {https://doi.org/10.1103/PhysRevApplied.15.024049} {\bibfield  {journal} {\bibinfo  {journal} {Phys. Rev. Applied}\ }\textbf {\bibinfo {volume} {15}},\ \bibinfo {pages} {024049} (\bibinfo {year} {2021})}\BibitemShut {NoStop}%
\bibitem [{\citenamefont {Brevoord}\ \emph {et~al.}(2025{\natexlab{b}})\citenamefont {Brevoord}, \citenamefont {Geus}, \citenamefont {Turan}, \citenamefont {Guerrero~Romero}, \citenamefont {Bedialauneta~Rodríguez}, \citenamefont {Codreanu}, \citenamefont {Stramma}, \citenamefont {Hanson}, \citenamefont {Elsen},\ and\ \citenamefont {Jungbluth}}]{brevoord_data_2025}%
  \BibitemOpen
  \bibfield  {author} {\bibinfo {author} {\bibfnamefont {J.~M.}\ \bibnamefont {Brevoord}}, \bibinfo {author} {\bibfnamefont {J.~F.}\ \bibnamefont {Geus}}, \bibinfo {author} {\bibfnamefont {T.}~\bibnamefont {Turan}}, \bibinfo {author} {\bibfnamefont {M.}~\bibnamefont {Guerrero~Romero}}, \bibinfo {author} {\bibfnamefont {D.}~\bibnamefont {Bedialauneta~Rodríguez}}, \bibinfo {author} {\bibfnamefont {N.}~\bibnamefont {Codreanu}}, \bibinfo {author} {\bibfnamefont {A.~M.}\ \bibnamefont {Stramma}}, \bibinfo {author} {\bibfnamefont {R.}~\bibnamefont {Hanson}}, \bibinfo {author} {\bibfnamefont {F.}~\bibnamefont {Elsen}},\ and\ \bibinfo {author} {\bibfnamefont {B.}~\bibnamefont {Jungbluth}},\ }\href {https://doi.org/10.4121/07dddb6a-8fcd-47a7-b8ad-584bafb8cfd5} {{\selectlanguage {english}\bibinfo {title} {Data underlying the publication "{Quantum} {Frequency} {Conversion} of {Single} {Photons} from a {Tin}-{Vacancy} {Center} in {Diamond}"}}} (\bibinfo {year} {2025}{\natexlab{b}})\BibitemShut {NoStop}%
\end{thebibliography}%

\end{document}